\documentclass[final]{agujournal2019} 
\usepackage{url} 
\usepackage{soul}
\usepackage{amssymb}
\draftfalse

\journalname{Journal of Geophysical Research: Machine Learning and Computation}

\begin{document}

%
%


\title{Deep Learning-Based Statistical Downscaling of Sea Surface Temperature Using a Residual Corrective Neural Network}

%
%




\authors{O. Jadhav\affil{1}, T. French\affil{2}, I. Janekovic\affil{3}, N. Jones\affil{1}, and M. Rayson\affil{1}}

\affiliation{1}{School of Earth and Oceans, UWA Oceans Institute, University of Western Australia, Crawley, WA, Australia.}
\affiliation{2}{School of Physics, Maths and Computing, Computer Science and Software Engineering, University of Western Australia, Crawley, WA, Australia.}
\affiliation{3}{School of Engineering, UWA Oceans Institute, University of Western Australia, Crawley, WA, Australia.}






\correspondingauthor{Onkar Jadhav}{onkar.jadhav@uwa.edu.au}



\begin{keypoints}
\item We present a residual corrective deep learning framework for downscaling the coarse sea surface temperature to high resolution.
\item The model captures fine-scale coastal features and marine heatwave anomalies absent in coarse climate data.
\item The model efficiently detects marine heatwave events even with few or no training examples. 
\end{keypoints}

%
%

%
%


\begin{abstract}
The large-scale oceanic and atmospheric forecasts provided by global climate models typically lack sufficient resolution to accurately capture the response of the coastal ocean to atmospheric forcing and coastal circulation that drive fine-scale SST variability. Dynamical downscaling is computationally prohibitive, when applied to extensive coastlines, predictive ensembles, or long time periods. Therefore, this work presents a statistical downscaling of sea surface temperature (SST) from the seasonal coupled ocean-atmosphere forecast system (ACCESS-S2) using machine learning techniques. This study proposes a novel deep learning framework that uses a U-Net to generate an initial high-resolution SST estimate, which is subsequently refined using a residual corrective approach. The target SST fields are derived from the Regional Ocean Modeling System (ROMS). This two step approach called Residual Corrective Neural Network (RCNN) progressively refines initial U-Net predictions by incorporating dynamically scaled residuals at each step, enabling accurate capture of broad patterns and fine-grained features such as eddies and fronts. We also introduce a custom loss-assisted RCNN variant to improve performance during extreme events, which may be absent from training data due to climate-driven shifts in SST extremes. The framework efficiently downscales SST along the west coast of Australia. A 2011 marine heatwave case study shows that the RCNN improves ACCESS-S2 SST predictions by increasing horizontal resolution from 25 km to 2 km, enabling identification of fine-scale anomalies unresolved in the ACCESS-S2 dataset. This balance between computational efficiency and accuracy supports applications in coastal impact assessment and marine ecosystem studies.
\end{abstract}

\section*{Plain Language Summary}
Global climate models are useful for understanding large-scale ocean and atmospheric patterns, but often lack the resolution needed to capture coastal changes in circulation and associated fine-scale SST variability, especially during extreme events such as marine heatwaves. Running high-resolution regional models that resolve these features is typically expensive and computationally intensive. This study introduces a two-step deep learning framework to enhance the spatial resolution of coastal sea surface temperature fields.

The first model predicts SST patterns, and the second, called a residual corrective neural network, refines those predictions to improve local detail. The framework improves coarse-resolution SST data (25 km) to high-resolution outputs (2 km), capturing fine-scale features like eddies and coastal fronts that are important for marine ecosystems.

In a case study of the 2011 marine heatwave off the west coast of Australia, the method successfully revealed small-scale temperature spikes and anomalies that were not visible in the global climate model data. This demonstrates the ability of the RCNN to detect extreme events even when few examples are present in the training data, by progressively correcting high temperature anomalies. Overall, this approach can help improve early warnings for ocean extremes and better support coastal planning and marine conservation.

%
%

%


%
%
%
%

\section{Introduction}
Sea surface temperature (SST) plays a critical role in driving oceanic and atmospheric processes \cite{deser2010}, influencing weather patterns \cite{trenberth2012}, marine ecosystems \cite{Ocarroll2019}, and climate extremes \cite{trenberth2012}. Accurate predictions of SST at high spatial resolution are essential for understanding and responding to regional-scale phenomena in a changing ocean such as marine heatwaves (MHWs), coastal upwelling, and biological productivity shifts \cite{chaudhary2023,oliver2021,patil2016}.
Various climate modeling systems, including global climate models (GCMs) and regional climate models (RCMs), offer SST forecasts at different spatial scales. GCMs such as Australian Community Climate and Earth-System Simulator – Seasonal (ACCESS-S) \cite{wedd2022access}, Community Earth System Model (CESM) \cite{danabasoglu2020community}, Max-Planck Institute Earth System model (MPI-ESM) \cite{mauritsen2019developments}, and European Centre for Medium-Range Weather Forecasts (ECMWF) \cite{molteni1996ecmwf} simulate large-scale ocean-atmosphere interactions and provide coarse-resolution SST fields useful for long-term global analyses. In contrast, RCMs like the Earth System Regional Climate Model (RegCM) \cite{elguindi2014regional}, and Regional Ocean Modeling System (ROMS) \cite{shchepetkin2005regional} operate at finer resolutions, capturing localized SST dynamics meso-scale features more effectively. The global ocean models typically have coarse horizontal spatial resolution ($\sim$25–50 km), making them insufficient for capturing fine-scale oceanic features, particularly in complex coastal regions \cite{tapiador2019climate,lupo2013global}.
Also, RCMs with a horizontal spatial resolution of 2 km are computationally prohibitive and therefore cannot be used to produce ensemble forecasts that assess the likelihood of extreme events, particularly when applied across large areas and longer time periods \cite{hobeichi2023using}.

Therefore, we developed a data-driven statistical downscaling  framework  to efficiently predict fine scale SST. This approach aims to learn the relationship between coarse-resolution GCM variables and fine-resolution regional variables from RCM. Here we also review applications to  precipitation and air temperature as machine learning–based SST downscaling has only been attempted in recent years.
Recent advances in machine learning (ML), especially deep learning, have enabled significant progress in statistical downscaling by capturing nonlinear and spatially complex relationships between predictors and targets.
A comprehensive review in \citeA{jebeile2021} highlights the rapid evolution of statistical downscaling methods, from early linear models to recent deep learning-based approaches. The authors of \citeA{Sachindra2018} applied machine learning models to downscale monthly precipitation under diverse climatic conditions. Likewise, \citeA{Vandal2019} compares traditional and machine learning-based statistical downscaling methods for daily and extreme precipitation. They compared baseline models like ordinary least squares and support vector regression, with neural networks. They find that advance neural network based approaches offer improvements in feature representation compared to simpler methods. \citeA{maqsood2023} compared random forest, support vector regression, and neural network models for statistically downscaling daily atmospheric temperature and precipitation over Prince Edward Island, finding that neural network models outperformed other methods. Furthermore, \citeA{hosseini2024} showed that for downscaling of atmospheric temperature and precipitation projections,  the statistical downscaling framework based on convolutional neural networks (CNN) outperformed fully connected feedforward neural networks, also referred to as multilayer perceptrons.

In short, the recent literature shows that deep learning models, particularly neural networks, have strong potential for spatial downscaling tasks due to their ability to represent complex spatial features \cite{Doury2023CD,Doury2024CD,Kendon2025BAMS}. Models like CNN, U-Net and autoenconders are also popular due to their ability to preserve hierarchical spatial features such as coastal fronts and eddies \cite{bano2021}.
Recent coastal downscaling studies have also demonstrated the effectiveness of CNN-based and deep learning super-resolution models for ocean variables, including Mediterranean SST super-resolution \cite{fanelli2024deep}, coastal water temperature downscaling in the northern Adriatic Sea \cite{adobbati2025deep}, coastal sea-state super-resolution \cite{kuehn2023deep}, and downscaling of sea surface height and currents in coastal regions \cite{yuan2024downscaling}.
Autoencoders are neural networks that learn compact representations of data by encoding and then reconstructing inputs, useful for tasks like dimensionality reduction. U-Nets extend this idea with an encoder-decoder structure and skip connections, making them especially effective for tasks where spatial detail is important, such as resolving sharp SST gradients during marine heatwaves.
However, standard CNN and U-Net models often suffer from oversmoothing and may fail to resolve localized structures, especially when trained on low-resolution inputs \cite{sharma2022resdeepd,vandal2017deepsd}.
To address this issue, \citeA{Rampal2024GRL,Rampal2025} introduced Conditional Generative Adversarial Networks for climate downscaling that can produce reliable and transferable high-resolution climate solutions. \citeA{vandal2017deepsd} introduced a stacked super-resolution convolutional neural network called DeepSD that focuses on progressively refining details. Moreover, \citeA{sharma2022resdeepd} proposed ResDeepD, a residual super-resolution network to downscale daily Indian monsoon precipitation using residual blocks. However, \citeA{Rampal2025} showed that some of these approaches lack explicit error correction mechanisms. \citeA{Mardani2025} proposed CorrDiff, a two-stage generative downscaling framework that combines deterministic UNet regression with a diffusion-based residual corrector to stochastically enhance km-scale atmospheric fields.
Generative models have also been explored for SST super-resolution and downscaling. For example, \cite{izumi2022super} and \cite{Kim2023} employed GAN-based approaches for SST super-resolution or data fusion, while \cite{Wang2024} proposed a diffusion-based SST downscaling model called DIFFDS. This approach provides a high-resolution version of the input SST and helps to restore most of the mesoscale processes. However, as with other data-driven generative models, diffusion-based downscaling models may face challenges under distribution shifts or rare extremes that are poorly represented in the training data, such as large SST increases associated with marine heatwaves in a warming ocean \cite{rampal2024extrapolation,Mardani2025}.

To summarize, there is a large body of literature investigating CNN-based approaches for statistical downscaling. However, a key limitation of standard CNN-based regression models is their tendency to regress toward the mean, which often results in overly smooth predictions and an inability to resolve fine-scale features such as coastal fronts and mesoscale eddies. Several studies have proposed generative approaches, including conditional generative adversarial networks, to overcome this limitation by introducing stochastic variability at smaller spatial scales. While effective, such approaches typically involve increased computational complexity and reduced interpretability.

Therefore, in contrast, our study proposes a residual-based refinement strategy that has not been widely explored for SST downscaling. We show that regression models, which are often perceived to “regress to the mean,” can in fact be substantially improved through an additional step that explicitly predicts and iteratively corrects residuals.
We present a residual corrective neural network (RCNN) framework that improves upon the traditional deep learning-based approach by introducing an iterative refinement process for downscaling sea surface temperatures. RCNN introduces structural modularity, interpretability, and targeted residual learning, making it better suited to both predict extreme climate events and achieve fine-scale SST reconstruction. In short, the modular structure of the RCNN makes it possible to adjust local features separately from the large-scale background SST, which provides a natural framework for handling different SST patterns. The RCNN first produces a preliminary high resolution prediction using a U-Net and then applies residual corrections in multiple steps to progressively improve spatial detail and accuracy. Furthermore, to enhance the ability of the model to predict extreme SST anomalies potentially not present in the training data, such as those observed during marine heatwaves, we introduce an extended version called custom loss-assisted RCNN (CL-RCNN), which incorporates synthetic MHW-like data and a custom loss function during training.
We generated synthetic data by incorporating physics-informed priors using Gaussian random fields. In addition, the developed approach is significantly faster and more energy efficient, making it more suitable for operational deployments or climate projection analysis.

The developed framework efficiently downscales the SST along the west coast of Australia. The model inputs included seven variables that were obtained from ERA5 and ACCESS-S2 models \cite{hersbach2020era5,wedd2022access}, while the target output was high-resolution SST from the Regional Ocean Modeling System (ROMS) \cite{shchepetkin2005regional}. All seven variables are used jointly as input predictors, while SST is the sole target variable downscaled in this study. 

This paper presents the architecture and implementation of the proposed RCNN and CL-RCNN frameworks in Section \ref{subsec:RCNN}. The performance of the framework is evaluated for both generalized SST prediction and for the 2011 Western Australian marine heatwave event in Subsections \ref{sec:generalCase} and \ref{sec:MHWresults}, respectively.
Additionally, we compare the RCNN approach with two independent models, namely, an interpolation baseline and a U-Net to enable a controlled comparison of downscaling performance across increasing model complexity.
The results demonstrate that the proposed models significantly outperform baseline approaches like interpolation, offering improved spatial fidelity and predictive accuracy, including under extreme conditions. Moreover, we studied how each input variable contributes in the final prediction in Subsection \ref{SHAPAnalysis} using Shapley Additive Explanations analysis (SHAP). Finally, generalization capabilities and potential overfitting behavior of the RCNN model are assessed in Subsection \ref{Overfitting}.

\section{Methods}
This section presents the proposed machine learning-based approach for downscaling sea surface temperature (SST) from coarse-resolution global ocean models ($\approx 25$ Km) to high-resolution regional scales ($\approx 2$ Km). The framework addresses the limitations of simple interpolation and introduces a deep learning method capable of preserving both large-scale spatial trends and fine-scale coastal features. 
First, Subsection \ref{subsec:RCNN} introduces the core model and methodology for the residual corrective neural network (RCNN) approach developed. Second, Subsection \ref{subsec:RCNN} addresses the challenge of predicting rare extreme events such as marine heatwaves, which are often underrepresented in the training data. Third, Subsection \ref{models} details the employed machine learning models. Finally, Subsection \ref{datasets} describes the datasets used as input and training data for the model. 

\subsection{Residual Corrective Neural Network}
\label{subsec:RCNN}
Let $X \in \mathbb{R}^{n\times r\times s\times p}$ be the input data, where $p$ represents the number of inputs gathered for a 2D spatial area of size $r\times s$ and $n$ is the number of samples. Similarly, the high-resolution output can be represented as $Y \in \mathbb{R}^{n\times h\times w\times q}$. Here, $h$ and $w$ denote the dimensions of the same 2D area used to extract inputs, but with higher resolution, i.e., $h > r, w> s$ and $q$ is the number of outputs.

To make the data suitable for regression, low-resolution inputs $X$ are mapped onto a finer grid of size $h\times w$ using an interpolation technique \cite{Du2020,li08,sekulic2020}. In this work, the low-resolution fields are interpolated to the high-resolution grid using a random-forest-based spatial interpolation method. This interpolation is performed independently for each predictor variable and each time step. For a given variable and time step, the interpolation model uses global spatial coordinates as inputs and the corresponding low-resolution variable values as outputs. Therefore, temporal variability is retained because the interpolation is repeated separately for each time step. This model learns a mapping $f:\mathbb{R}^2 \rightarrow \mathbb{R}$ from spatial coordinates (latitude and longitude) to the values of low-resolution variables \cite{Hengl2018}. Given a pair of coordinates, the trained model produces an estimated value for the variable at that location, effectively performing the interpolation. Therefore, when the fine-grid spatial coordinates are fed to the trained model, it interpolates the low-resolution variables onto the high-resolution grid. However, simply interpolating the low-resolution inputs on to a high-resolution grid only reshapes the data it does not add any new information therefore it cannot resolve the fine-scale processes. Therefore, a statistical downscaling framework is necessary to establish a relationship between these interpolated low-resolution global climate variables and fine-scale local climate variables \cite{jebeile2021,maraun2018,sun2024}. In this work, the statistical downscaling framework has been developed using deep learning models.

This interpolation step only maps the low-resolution predictors onto the high-resolution grid and does not add new fine-scale physical information. It is therefore used both as a preprocessing step for the learning models and as an interpolation baseline against which the downscaling models are compared.

Convolutional neural networks are widely used for statistical downscaling because they can learn complex spatial relationships between coarse-scale predictors and fine-scale targets \cite{Goodfellow2016,ronneberger2015,vasilev2019}. However, when the input information is derived from coarse-resolution global models, standard CNNs often produce overly smooth outputs and struggle to recover sharp coastal gradients or small-scale features. This limitation arises because the interpolated low-resolution inputs mainly describe the large-scale background state and contain little direct information about fine-scale variability. To address this, we adopt an iterative residual correction approach, where the model first predicts a high-resolution SST field and then progressively refines it through targeted corrections. In this framework, the inputs are coarse-resolution variables from global climate models, while the high-resolution SST fields from the regional ROMS model are used as the training target.

The RCNN combines U-Net as an initial high-resolution prediction model with a residual corrective network to iteratively refine neural network predictions, guiding them toward high-resolution targets. In short, the initial high-resolution prediction is generated by the U-Net and serves as the starting point for residual refinement. This approach progressively corrects the initial high-resolution predictions through learned residuals, making it particularly suitable for spatial data requiring high fidelity. This approach aims to balance global structure and local detail for enhanced high-resolution predictions.

In short, instead of directly predicting the high-resolution solution, $y$, the RCNN model breaks the approach into two steps. In the first step, a U-Net predicts an initial high-resolution solution $y_{\mathrm{initial}}$, which is a close estimate of the high-resolution solution, $y$.
Assuming that $y_{\mathrm{initial}}$ is a reasonably accurate approximation of $y$, the residual $r = y - y_{\mathrm{initial}}$ should  exhibit less variance than the original high-resolution target $y$ \cite{liitiainen2009residual}. With this in mind, training the model on a simplified, lower-variance target makes the task easier and tractable. Therefore, in the second step of RCNN, a new model is trained with this low-variance residual $r$ as a target. A straightforward approach would involve using a simpler U-Net-type model to predict this residual $r$ and correct the solution $y_{\mathrm{initial}}$ by adding this predicted residual $r$. However, while this method may appear simple, it introduces significant challenges. Predicting the residual in one step requires the model to learn all spatial correlations simultaneously. This can lead to poor results because the high-resolution SST field exhibits both broad spatial trends and localized features, such as eddies or fronts. A single step residual prediction model has to learn these complex patterns in a single pass, which may not be sufficient to capture the fine details. 

In contrast, the RCNN approach does not correct the initial prediction in a single step. By breaking the task into smaller, more manageable steps, the RCNN ensures that the model learns spatial correlations incrementally, improving its ability to capture both broad trends and localized features.

\begin{figure}[htb]
  \centering
  \includegraphics[width=1\columnwidth]{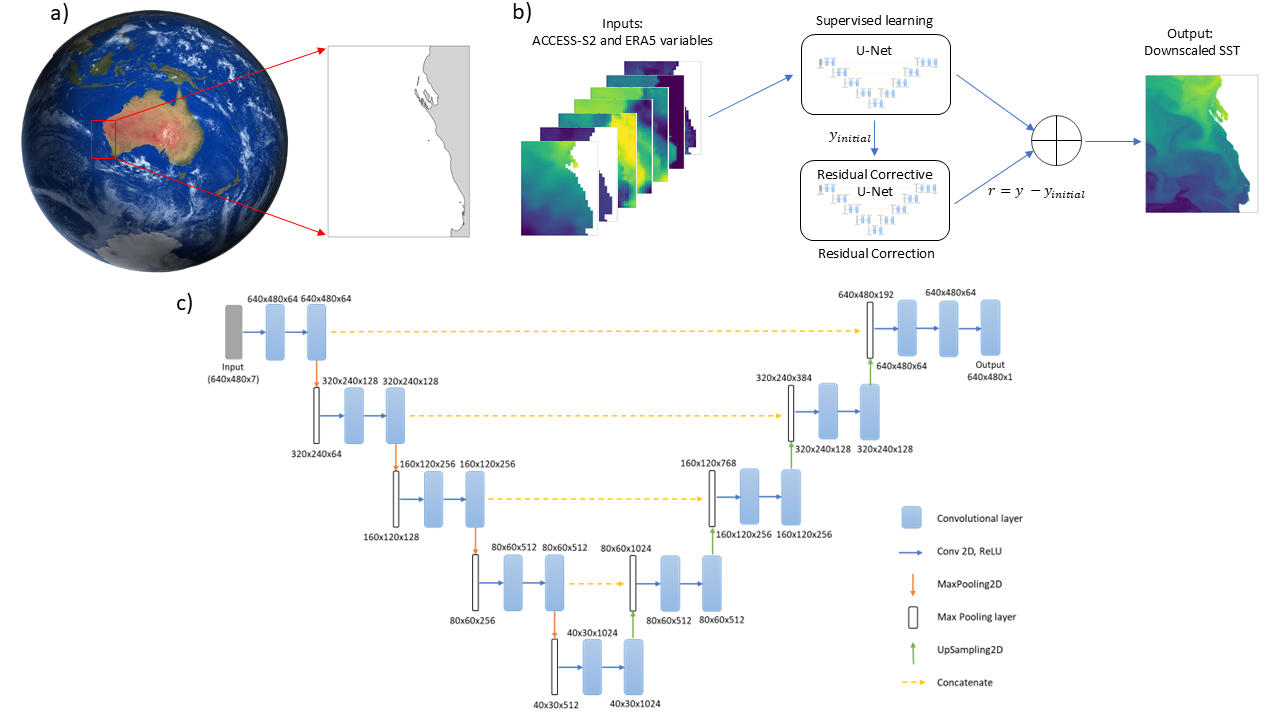}
  \caption{Overview of the Residual Correction Neural Network and the U-Net-based framework developed for statistical downscaling. (a) The target region of interest along the west coast of Australia (b) A workflow depicting the developed residual-corrective approach. The RCNN framework employs a U-Net to predict an initial rough solution $y_{\mathrm{initial}}$, which is then corrected using the residual corrective network trained on $r= y-y_{\mathrm{initial}}$ to achieve the high-resolution SST field $y$. The U-Net provides only the initial prediction; residual refinement is performed by the RCNN. (c) Detailed U-Net architecture used for residual prediction, consisting of convolutional layers, pooling layers, and upsampling layers with skip connections.}
  \label{fig:1}
\end{figure}

The process begins by setting the initial prediction $y_{\mathrm{initial}}$ as the initial guess of the high-resolution solution
\begin{linenomath*}
\[
    \hat{y}_T = y_{\mathrm{initial}}.
\]
\end{linenomath*}
For each iteration step $t \in \{T, T-1, \dots, 1\}$, the model refines this coarse prediction to obtain a high-resolution prediction that is closer to the actual truth $y$. The refinement process is defined as
\begin{linenomath*}
\begin{equation}
    \hat{y}_{t-1} = \hat{y}_t + \alpha_t \Delta \hat{y}_t,
    \label{eq:1}
\end{equation}
\end{linenomath*}
where $\Delta \hat{y}_t$ is a corrective term predicted by the residual correction network. During training, this corrective term is learned to approximate the residual between the high-resolution ROMS target $y$ and the current prediction $\hat{y}_t$, i.e.,
\begin{linenomath*}
\[
    \Delta \hat{y}_t \approx  y - \hat{y}_t.
\]
\end{linenomath*}
Thus, the corrective term $\Delta \hat{y}_t$ in equation (\ref{eq:1}) is progressively added to improve the solution by multiplying it by a scaling factor $\alpha_t$. Due to this iterative refinement, the model shifts its focus from coarse corrections in the early steps to fine-grained refinements in the later steps. In this work, the corrective term $\Delta \hat{y}_t$ is predicted by a simpler U-Net (defined in Fig. \ref{fig:1}). This neural network is trained once. During training, it takes $y_{\mathrm{initial}}$ and the interpolated low-resolution SST field $X_{\mathrm{SST}}^{\mathrm{interp}}$ as input channels, and learns the residual target $r = y - y_{\mathrm{initial}}$. During recursive refinement, the same trained network is reused by replacing $y_{\mathrm{initial}}$ with the current prediction $\hat{y}_t$, while keeping $X_{\mathrm{SST}}^{\mathrm{interp}}$ fixed as an input. The ROMS target $y$ is used only to define the residual target during training and is not used as an input to the residual correction network.
Therefore, for the initial residual prediction, the network approximates
\begin{linenomath*}
\[
    \Delta \hat{y}_{T} =
    f_\theta(y_{\mathrm{initial}}, X_{\mathrm{SST}}^{\mathrm{interp}})
    \approx y - y_{\mathrm{initial}}.
\]
\end{linenomath*}
indicating that the new U-Net, $f_\theta$, learns to predict the corrective term $\Delta \hat{y}_t$ that approximates the residual $r$.
In the subsequent steps $t=T-1, T-2, \dots, 1$, this trained network is used to predict the next corrective term from the current prediction and the fixed interpolated SST input:
\begin{linenomath*}
\[
    \Delta \hat{y}_{t} =
    f_\theta(\hat{y}_{t}, X_{\mathrm{SST}}^{\mathrm{interp}}).
\]
\end{linenomath*}

We tested different schedules for $\alpha$ such as constant, linear decay schemes, and cosine. We found that constant $\alpha$ led to over-correction in later iterations or linear decay caused insufficient early correction followed by premature stagnation. In contrast, the cosine schedule provided smoother convergence and more stable refinement across iterations, preventing error amplification while preserving fine-scale structure. This behavior aligns with the objective of decomposing the correction process into manageable, progressively finer updates. Importantly, the choice of $\alpha$ only controls the step size of each correction and does not constrain the magnitude of the learned residual itself.
Therefore, the alpha schedule for $T$ iterations is defined as
\begin{linenomath*}
\[
    \alpha_t =
    \alpha_{\mathrm{min}} +
    \frac{1}{2}(\alpha_{\mathrm{max}} - \alpha_{\mathrm{min}})
    \left(1 - \mathrm{cos}\left(\frac{t \pi}{T}\right)\right),
\]
\end{linenomath*}
where $\alpha_{\mathrm{max}}, \alpha_{\mathrm{min}}$ define the upper and lower bounds of the correction factor. With this reverse indexing, $\alpha_t$ is larger near the initial refinement step and smaller near the final refinement step. This cosine schedule governs the smooth refinement over $T$ iterations. 
Here, $\alpha_{\min}$ and $\alpha_{\max}$ were selected through randomized search on the held-out validation set. In other applications, these parameters and the number of correction steps $T$ can be tuned similarly using validation RMSE and SSIM while keeping the test period completely unseen.

Therefore, after $T$ iterations of the refinement, the final prediction $\hat{y}_0$ is
\begin{linenomath*}
\[
    \hat{y}_0 = \hat{y}_T + \sum_{t = 1}^T ( \alpha_t \Delta \hat{y}_t).
\]
\end{linenomath*}
Note that, at each $t^{\mathrm{th}}$ iteration, the model is not retrained. This is because the model trained on the initial residual $r$ has already learned a generalized function to predict corrections $\Delta \hat{y}_t $ for a given input. This function  works well for the intermediate solutions during refinement, as they are still within the distribution of residuals seen during training. 

In addition, an iterative approach may sometimes accumulate errors instead of refining the solution, especially if the residual correction is noisy or incorrectly estimated. To mitigate this issue, an adaptive scaling factor is introduced, which verifies the effectiveness of each correction step before applying it. This is done through a residual verification step, ensuring that each correction  $\Delta \hat{y}_t$ actually improves the solution before being added iteratively.
Note that the recursive refinement is evaluated on the validation set. The quantities $\ell_t$ and $\ell_{t-1}$ are regularized validation errors used to assess the effect of each recursive correction step as follows. They are not used to retrain the residual network at each iteration.
As a first step, before adding the corrective term, the regularized validation error between the solution $\hat{y}_{t}$ and the true solution $y$ is computed
\begin{linenomath*}
\begin{equation}
   \ell_{t} =
   \ell_{\mathrm{MSE}} \big ({\hat{y}_t, y} \big )
   + \lambda \| \Delta \hat{y}_t \|_2.
   \label{eq:2}
\end{equation}
\end{linenomath*}
The regularization term penalizes excessively large corrective updates and helps prevent error amplification during recursive refinement.

In the second step, a new regularized validation error is computed by adding the corrective term
\begin{linenomath*}
\[
    \ell_{t-1} = \ell_{\mathrm{MSE}} \big ({\hat{y}_t + \alpha_t \Delta \hat{y}_t, y} \big ) + \lambda \| \Delta \hat{y}_{t} \|_2.
\]
\end{linenomath*}
If $\ell_{t-1} < \ell_t$, then the correction is accepted as it improves the regularized validation error. Otherwise, the scaling factor $\alpha_t$ is reduced as
\begin{linenomath*}
\[
    \alpha_t = \alpha_t \times \mathrm{max}\bigg ( \frac{\ell_t - \ell_{t-1}}{\ell_{t}}, 0.5 \bigg).
\]
\end{linenomath*}
This ensures that if the correction increases the error, it receives a smaller weighting.

Furthermore, the refinement process terminates once the specified number of iterations $T$ is completed, ensuring computational efficiency by avoiding unnecessary iterations. Figure \ref{fig:1} shows a generalized framework of this approach. In this study, recursive refinement was evaluated on the validation set, and a maximum of $T=20$ correction steps was used for all reported experiments.

Note that, to ensure accurate predictions and prevent model bias, we incorporated a masking mechanism that restricts computations to valid ocean regions and excludes land areas. Although common preprocessing techniques, such as replacing NaNs with mean or maximum values, can provide complete input tensors, they risk introducing artificial signals or distorting physical patterns, especially around land-ocean boundaries. Such an imputation may also mislead the model into learning from irrelevant data points. Instead, we adopt a more physically consistent approach by applying a binary land-ocean mask during the prediction stage. This mask is a fixed 2D array with ones indicating valid ocean pixels and zeros representing land regions. The convolutional layers do not explicitly skip masked regions internally; standard convolutions are applied to complete gridded tensors. Before training and inference, NaN values over land are replaced by a neutral fill value and the input and target fields are multiplied by the binary land-ocean mask. The same mask is applied in the loss calculation, so that only valid ocean pixels contribute to optimization and evaluation. Thus, land pixels are retained only as neutral placeholders required for convolutional operations and do not affect the learned objective.

\subsubsection*{Improvements for Marine Heatwave Prediction}
\label{ImprovementMHW}
The approach defined above will work  well for  cases where the input SST range in the forecast matches the SST range of the training data. However, this model may fail to predict marine heat wave events, where  regions experience unusually high and unprecedented SST values. This limitation arises because the training data set used to train the model may not have contained such high temperatures. As a result, the model might struggle to approximate unprecedented temperatures accurately, leading to underpredicted SST. 

To overcome this issue, model retraining with synthetic SST data that simulates MHW-like events is employed. The synthetic data generation is later explained in this subsection. This retraining step aims to improve the ability of the model to predict extreme SST conditions. 
The key component of this approach is the addition of a custom loss function to guide the model during retraining. There are  two key objectives. First, the model must continue to accurately predict the standard temperature patterns. This can be achieved by using a standard mean squared error loss between the model prediction and true SST field, given by (\ref{eq:2}). Second, the model should also learn to replicate the MHW events introduced by the synthetic data. To enforce this, an additional MSE term is added to the original MSE. The final custom loss reads
\begin{linenomath*}
\begin{equation}
    \ell_t =
    \ell_{\mathrm{MSE}}(\hat{y}_t, y)
    + \beta \ell_{\mathrm{MSE}}(\hat{y}_t, y_{\mathrm{synthetic}})
    + \lambda \| \Delta \hat{y}_t \|_2,
    \label{eq:3}
\end{equation}
\end{linenomath*}
where the first term of (\ref{eq:3}) represents the MSE between the model prediction and the ROMS target, while the second term represents the MSE between the model prediction and the synthetic MHW-like target. The third term is the same correction regularization used in the RCNN formulation and penalizes excessively large residual updates. Additionally, $\beta$ is a hyperparameter that controls the model learning based on the synthetic data. It is a balancing factor between true and synthetic loss that is critical for controlling how much the model follows the synthetic data without compromising its general accuracy. Also, the additional MSE term encourages the model to learn and reproduce high-temperature values from the synthetic data, which are crucial to forecasting future MHW events.
In short, this simple retraining process enables the model to predict high temperature values while maintaining its ability to predict the standard temperature range. 
This custom loss-assisted RCNN (CL-RCNN) uses the same inference inputs and residual correction architecture as the standard RCNN. In particular, the network input during recursive refinement remains the current prediction $\hat{y}_t$ together with the interpolated low-resolution SST field $X_{\mathrm{SST}}^{\mathrm{interp}}$. The ROMS field $y$ and the synthetic field $y_{\mathrm{synthetic}}$ are not inputs to the network. They are used only as target fields in the training loss.

In this work, synthetic high-resolution SST fields are generated by adding a scaled Gaussian Random Field (GRF) to the interpolated low-resolution SST fields.
We generate a physically realistic, high-resolution SST field that captures extreme marine heatwave signals by injecting modulated fine-scale Gaussian features into a reference ROMS SST field. This approach combines the spatial structure of high-resolution model output with large-scale thermal anomalies derived from lower-resolution interpolated MHW data.
Let $y_{\mathrm{LR}}$ denote the low resolution interpolated SST field for a MHW-like event. We isolate the large-scale SST anomalies by removing a spatially smoothed background as
\begin{linenomath*}
\begin{equation}
    \Delta_{\mathrm{MHW}} = y_{\mathrm{LR}} - G_{\sigma_G} (y_{\mathrm{LR}}),
    \label{eq:4}
\end{equation}
\end{linenomath*}
where where $G_{\sigma_G}$ is a Gaussian smoothing operator with standard deviation $\sigma_G$. 
Let $y$ be defined as some high-resolution SST field available in the training dataset that can be modified to emulate a MHW event. We generate a Gaussian random field whose power spectrum matches that of $y$ as
\begin{linenomath*}
\[
    \Phi = \mathrm{Re}[F^{-1}(|F(y)|\cdot e^{i\theta})],
\]
\end{linenomath*}
where $F$ and $F^{-1}$ denote the 2D Fourier and inverse Fourier transforms. $|F(y)|$ is the amplitude spectrum of the field $y$.
The extracted anomaly from (\ref{eq:4}) is then modulated by
\begin{linenomath*}
\[
    \Delta = \Phi \odot \Delta_{\mathrm{MHW}},
\]
\end{linenomath*}
where $\odot$ is the Hadamard product. The final synthetic high-resolution SST field is constructed by summing the modulated anomaly with the original $y$ 
\begin{linenomath*}
\[
    y_{\mathrm{synthetic}} = y + \gamma\Delta,
\]
\end{linenomath*}
where $\gamma$ is the scaling factor that controls the strength of synthetic injection. The scaling factor $\gamma$ is treated as a sensitivity parameter rather than a physically prescribed quantity. Its value is selected empirically through trial-and-error testing on validation data.

Similar methods of generating synthetic fine-scale SST variability by randomizing Fourier phases while preserving the spatial power spectrum follow standard approaches used in geostatistics and climate science for simulating spatial fields. A more detailed discussion on synthetic data generation using GRF can be found in \cite{kleiber2016high,l2019gaussian,vaccaro2021climate}.  
It is important to note that the synthetic SST generation assumes that mesoscale spatial variability retains broadly similar statistical characteristics during marine heatwaves. In particular, we assume that the statistical properties of mesoscale oceanographic features, such as characteristic eddy scales and frontal gradients, remain approximately preserved under extreme thermal conditions. This assumption is physically supported by the scale separation between mesoscale dynamics (10-200 km, days to weeks) and the development of basin-scale marine heatwaves (months to seasons), as well as by observational evidence from the 2011 Ningaloo Nino event, which exhibited persistent mesoscale structures despite extreme SST anomalies \cite{feng2013nina}. Preserving the spatial power spectrum of high-resolution ROMS fields while modulating anomaly amplitude therefore provides a physically grounded first-order approximation for representing extreme SST states. The validation of our downscaled ensemble during the 2011 MHW event (Section \ref{sec:MHWresults}) provides empirical support for this assumption within the observed range of extremes. We also note that this assumption may be less valid in regions or scenarios where circulation dynamics undergo substantial structural reorganization.

In short, this study evaluates three models. (i) A standalone U-Net is used as a baseline to predict an initial high-resolution SST field. (ii) The RCNN refines this U-Net output through iterative residual correction using a lightweight corrective network. (iii) The CL-RCNN retains the same architecture and inference inputs as RCNN but is retrained using synthetic marine heatwave target fields and a custom regularized loss function to improve performance on rare extreme events.

\subsection{Machine Learning Models}
\label{models}
This subsection presents the machine learning models used for spatial downscaling, including both traditional and deep learning approaches. It explains the baseline model used for interpolation as well as the fully convolutional U-Net architecture designed for high-resolution SST reconstruction.

We use a random forest regression model as the baseline spatial interpolation model, where the regressor learns a mapping from spatial coordinates to the corresponding predictor-variable values.
It is an ensemble learning algorithm that builds multiple decision trees and averages their predictions for better generalization \cite{breiman2001random,sekulic2020random}. The model takes a pair of coordinates and estimates the corresponding predictor-variable value at that location. Specifically, for each predictor variable and time step, the latitude and longitude of each valid low-resolution grid point are used as the two input features, while the corresponding variable value is used as the regression target; the trained model is then evaluated at the high-resolution ROMS-grid coordinates. The random forest regressor used in this study consists of 500 decision trees, chosen to balance accuracy and computational efficiency. The maximum depth of each tree was left unrestricted, allowing the trees to grow fully. The splits within each tree were determined with a minimum of two samples per split. Bootstrapping was enabled to ensure robust tree diversity and a random state of 42 was used for reproducibility. We used the scikit-learn implementation of random forest \cite{scikitlearn}. Because the random forest was fitted only to low-resolution coordinate–value pairs, no separate cross-validation against ROMS was performed; instead, its interpolated outputs were evaluated against ROMS over the same independent test periods as the U-Net and RCNN. The seven input variables described in Tab. \ref{Tab:Features} of the model were standardized before training to have zero mean and unit variance. 

\begin{table}[htb]
\caption[U-Net with masking]{%
Architecture of the U-Net model with masking. All Conv2D layers use He-normal initialization and are followed by Batch Normalization. A Dropout layer is applied at the bottleneck with dropout rate of 0.2.
}
\label{Tab:UNet}
\begin{center}
\resizebox{\textwidth}{!}{
\begin{tabular}{ l c c c c l }
 \hline
Layer Name         & Type         & Filters & Kernel Size & Activation & Additional Details \\
 \hline
Input (Main Data)   & Input       & –   & (h, w, 7) & –    & Main input (Tab.~\ref{Tab:Features}) \\
Input (Mask)        & Input       & –   & (h, w, 1) & –    & Binary land–ocean mask \\
Conv Block 1        & Conv2D      & 64  & 3×3       & ReLU & – \\
Conv Block 1        & Conv2D      & 64  & 3×3       & ReLU & – \\
Pooling 1           & MaxPooling2D& –   & 2×2       & –    & – \\
Conv Block 2        & Conv2D      & 128 & 3×3       & ReLU & – \\
Conv Block 2        & Conv2D      & 128 & 3×3       & ReLU & – \\
Pooling 2           & MaxPooling2D& –   & 2×2       & –    & – \\
Conv Block 3        & Conv2D      & 256 & 3×3       & ReLU & – \\
Conv Block 3        & Conv2D      & 256 & 3×3       & ReLU & – \\
Pooling 3           & MaxPooling2D& –   & 2×2       & –    & – \\
Bottleneck          & Conv2D      & 512 & 3×3       & ReLU & Deepest feature representation \\
Upsampling 1        & UpSampling2D& –   & 2×2       & –    & – \\
Concatenate 1       & Concatenate & –   & –         & –    & Skip connection with Conv Block 3 \\
Conv Block 4        & Conv2D      & 256 & 3×3       & ReLU & – \\
Conv Block 4        & Conv2D      & 256 & 3×3       & ReLU & – \\
Upsampling 2        & UpSampling2D& –   & 2×2       & –    & – \\
Concatenate 2       & Concatenate & –   & –         & –    & Skip connection with Conv Block 2 \\
Conv Block 5        & Conv2D      & 128 & 3×3       & ReLU & – \\
Conv Block 5        & Conv2D      & 128 & 3×3       & ReLU & – \\
Upsampling 3        & UpSampling2D& –   & 2×2       & –    & – \\
Concatenate 3       & Concatenate & –   & –         & –    & Skip connection with Conv Block 1 \\
Conv Block 6        & Conv2D      & 64  & 3×3       & ReLU & – \\
Conv Block 6        & Conv2D      & 64  & 3×3       & ReLU & – \\
Output Layer        & Conv2D      & 1   & 1×1       & Linear & Generates high-resolution SST \\
Masking             & Multiply    & –   & –         & –    & Applies land-ocean mask \\
\hline
\end{tabular}
}
\end{center}
\end{table}

The second model used for the downscaling task is a convolutional neural network called U-Net \cite{iglovikov2018ternausnet,prince2023understanding,zhang2023dive}. It follows an encoder-decoder architecture, where the encoder extracts hierarchical features from the interpolated inputs, and the decoder reconstructs a high-resolution output while preserving spatial details through skip connections. Skip connections directly transfer feature maps from the encoder to the decoder, preserving spatial detail that would otherwise be lost during downsampling. The model takes a multi-channel input, where each channel corresponds to an input variable presented in Tab. \ref{Tab:Features}. Similar type of U-Nets have been used for downscaling before \cite{Meer2023}

The encoder has three downsampling blocks, each containing two 3×3 convolutional layers with a ReLU activation function. It is followed by max-pooling (2×2) to sequentially extract hierarchical features \cite{sharma2017activation}. 
The bottleneck layer with 512 filters forms the lowest-dimensional latent representation of the input, capturing the most abstract features before the decoder stage reconstructs the fine-resolution output.
The decoder has a similar structure that uses UpSampling2D layers to restore spatial resolution. Each upsampled feature map is concatenated with the corresponding encoder feature map via skip connections, helping to retain fine-scale details lost during downsampling. The final output is an 1x1 convolution layer with a linear activation function to obtain the high-resolution SST field. Additionally, this model incorporates a masking mechanism, which ensures that predictions are only computed over valid ocean regions, excluding land pixels. This is achieved by multiplying the final SST output with a predefined binary land-ocean mask.

The model is trained using mean squared error (MSE) loss, optimized with the Adamax optimizer, and uses normalized input features to stabilize learning.
All neural-network models were trained using Adamax with an initial learning rate of $10^{-3}$, a batch size of 8, and a maximum of 50 epochs. 
The checkpoint with the lowest validation RMSE was retained. The MSE is given as
\begin{linenomath*}
\[
    \ell_{\mathrm{MSE}}(\hat{y}, y) = \frac{1}{n}\sum_{i=1}^n (y_i - \hat{y}_i)^2
\]
\end{linenomath*}
No separate early-stopping criterion was used.

The detailed structure of the defined U-Net can be found in Tab. \ref{Tab:UNet}.

The U-Net based RCNN model uses the same U-Net structure defined above for the initial high-resolution solution $y_{\mathrm{initial}}$. However, for the residual learning, a smaller, more computationally efficient U-Net has been used. This version reduces the number of filters in each layer and decreases the model depth while maintaining the same encoder-decoder architecture and skip connections. The number of convolutional filters were reduced to a maximum of 256 instead of 512. Also, there are only 3 downsampling layers instead of 4. The rest of the structure remains similar to the main U-Net. Also, the values of $\alpha_{\mathrm{max}} = 0.65$ and $\alpha_{\mathrm{min}} = 0.15$ are used in the cosine function. 

Note that the training of the models was conducted on a high-performance computing environment utilizing a single GPU allocation on the Setonix system. Each allocation consists of 1 AMD Instinct MI250X GPU, 8 CPU cores (1 chiplet), and approximately 29.44 GB of system RAM. 

\subsection{Datasets}
\label{datasets}
To train these ML models, we used three different datasets. The inputs/Predictor variables for the model are gathered from the ERA5 \cite{hersbach2020era5} and ACCESS-S2 \cite{wedd2022access} reanalysis datasets. The target SSTs are obtained from the ROMS model. There are seven inputs as shown in Tab. \ref{Tab:Features}: low-resolution sea surface temperature, salinity, and mixed layer depth from ACCESS-S2, along with surface latent heat flux, surface net solar radiation, surface net thermal radiation, and surface sensible heat flux from ERA5. 
\begin{figure}[htb]
  \centering
  \includegraphics[width=0.7\columnwidth]{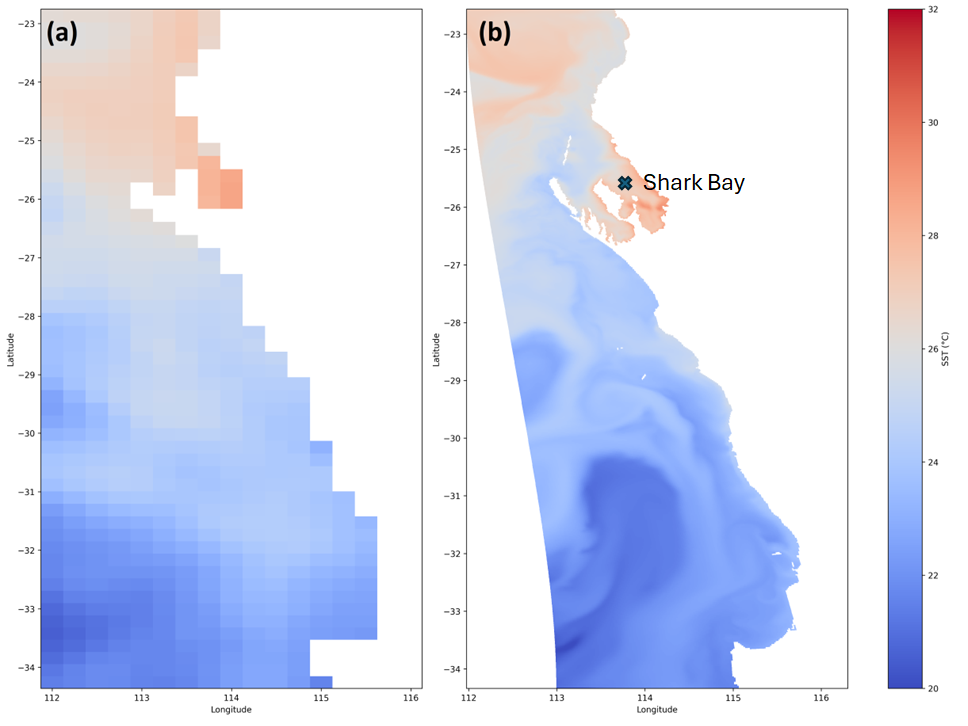}
  \caption{Comparison of SST fields from (a) ACCESS-S2 and (b) the high-resolution ROMS model for the same region off the coast of Western Australia on the 31st Jan 2021. We have also marked the Shark Bay coastal location (25.78$^\circ$S, 113.29$^\circ$E) used in this study for further analysis.}
  \label{fig:GlobalvsLocal}
\end{figure}
The model output - the target vector – consisted of high-resolution sea surface temperature fields derived from a regional 3D numerical model.
Specifically, we used hourly data spanning 22 years (2000–2021) from a high resolution ocean reanalysis focused on the central west coast of Australia. Note that, all predictor variables are used jointly as multichannel inputs to the downscaling models, while SST is the sole predictand; auxiliary variables are not independently downscaled. The ERA5 and ACCESS-S2 predictors were not bias-corrected; they were interpolated onto the ROMS grid and standardized using training-set statistics before model input.

The reanalysis employed a fine-scale configuration ($\sim$ 2 km horizontal resolution with 25 vertical layers, including enhanced surface resolution) of the three-dimensional, nonlinear, free-surface, s-coordinate primitive equation Regional Ocean Modeling System \cite{shchepetkin2009correction}.
Initial and open boundary conditions for the local model were sourced from Mercator Ocean International’s operational analysis and forecasting system, which provides data at a 1/12$^\circ$ horizontal resolution. Atmospheric forcing applied at the surface boundary was used to provide variables for momentum, fresh water and heat fluxes and was based on the ERA5 reanalysis from the European Centre for Medium-Range Weather Forecasts (ECMWF), with 25 km horizontal resolution and hourly temporal resolution. Additionally, tidal forcing was incorporated using eight principal tidal constituents from the TPXO9 global tidal solution \cite{egbert2002efficient}. This modelling setup has been previously been applied in similar coastal applications \cite{mahjabin2019spatial,rafiq2020dynamics,van_der_mheen_substantial_2024}.
\begin{table}[htb]
\caption{All considered inputs for the machine learning model. Sea surface temperature (SST), salinity (salt), and mixed layer depth (mld) are obtained from ACCESS-S2, while surface latent heat flux (slhf), surface net solar radiation (ssr), surface net thermal (longwave) radiation (str), and surface sensible heat flux (sshf) are obtained from ERA5.}
\label{Tab:Features}
\begin{center}
\begin{tabular}{ l c }
 \hline
ACCESS-S2 & Abbreviation\\
 \hline
Sea surface temperature & SST \\
\noalign{\smallskip}
Salinity & salt \\
\noalign{\smallskip}
Mixed layer depth & mld\\
 \hline
ERA5 & \\
 \hline
Surface latent heat flux & slhf\\
\noalign{\smallskip}
Surface net solar radiation & ssr\\
\noalign{\smallskip}
Surface net thermal (longwave) radiation & str\\
\noalign{\smallskip}
Surface sensible heat flux & sshf\\
\hline
\end{tabular}
\end{center}
\end{table}
The original GCM datasets, ACCESS-S2 and ERA5, provide global coverage with spatial resolutions of approximately 25 km and 25 km, respectively. However, for this study, we focus on a specific region along the west coast of Australia, defined by longitudes between 108 $^\circ$E and 117$^\circ$E and latitudes from 34$^\circ$S to 21$^\circ$S. To ensure consistency across datasets, we extract ROMS data with a spatial resolution of approximately 2 km for the same geographic region. We emphasize that this study is exclusively concerned with SST, and all land-based grid points have been excluded from the analysis. Also, all predictor variables are provided simultaneously as multi-channel inputs to the model, allowing it to learn their joint covariance with SST rather than treating them independently. The ML datasets used span a period from 2000 to 2021, providing a comprehensive temporal coverage for investigating long-term SST variations and potential trends in the region.

Figure \ref{fig:GlobalvsLocal} shows the comparison between the SST fields from ACCESS-S2 and ROMS. The ACCESS-S2 SST (Fig. \ref{fig:GlobalvsLocal}(a)) is displayed on a coarse rectilinear grid and has a blocky appearance that lacks fine-scale features. The grid resolution cannot resolve submesoscale processes, creates diffuse representations of thermal gradients across mesoscale features and  leads to a simplified representation of the coastline. In contrast, the ROMS SST field (Fig. \ref{fig:GlobalvsLocal}(b)) is based on a curvilinear grid with significantly higher spatial resolution, allowing it to capture coastal geometry, eddy structures, and thermal gradients with much greater fidelity.

Furthermore, the selection of an appropriate training dataset plays an important role in model performance. In particular, the temporal span and characteristics of the training data can significantly affect the model’s capacity to learn relevant patterns, especially in geophysical applications where temporal variability and non-stationarity are inherent. To explore this, we conducted a comprehensive analysis by systematically varying the time windows used for model training.
We formed multiple training datasets by selecting different temporal subsets of the available data, each representing distinct periods with varying climatic or environmental conditions. These training windows differed in both their starting points and durations to capture the impact of short-term vs. long-term training data and the inclusion or exclusion of extreme events or transitions. For each configuration, a separate RCNN model was trained while keeping all other parameters, architecture, and preprocessing steps constant. This setup allowed us to isolate the effect of the selection of training data on the performance of the model. The model evaluations were then performed on a fixed test dataset to ensure comparability. 

Through this analysis, our objective was to determine the sensitivity of the RCNN framework to variations in training data and to identify training windows that produce the most stable and accurate predictions.
For the generalized experiments, the final three months of each temporal configuration were reserved for validation and excluded from model fitting. For example, the 2015–2020 configuration contained 2100 training samples from January 2015 to September 2020 and 92 validation samples from October to December 2020, followed by independent evaluation on 365 daily samples from 2021. No spatial holdout was used; the data separation was strictly temporal. 
For the marine-heatwave experiment, 2099 daily samples from January 2005
to September 2010 were used for model fitting, 92 samples from
October--December 2010 were reserved for validation, and the 90 daily
ROMS fields from January--March 2011 were withheld for evaluation.
No spatial holdout was used.

\subsection{Metrics}
In this work, the models presented in subsection \ref{models} are evaluated on various quantitative performance metrics, such as Root Mean Square Error (RMSE) \cite{hodson2022root}, Structural Similarity Index (SSIM) \cite{brunet2011mathematical}, and the coefficient of determination $R^2$ \cite{di2008coefficient}. 

The RMSE measures the average magnitude of the prediction errors between the true $y$ and predicted values $\hat{y}$, and is given as
\begin{linenomath*}
\[
    \mathrm{RMSE}(\hat{y}, y) = \sqrt{\frac{1}{n}\sum_{i=1}^n (y_i - \hat{y}_i)^2}.
\]
\end{linenomath*}
RMSE is sensitive to large errors due to the squared term, making it especially informative when outliers or large deviations are of concern. Lower RMSE values indicate better model performance in terms of raw accuracy. 

SSIM is a perceptual metric designed to evaluate the similarity between two images or 2D spatial fields. Unlike traditional pixel-wise error measures such as RMSE, SSIM captures structural information. SSIM captures how well the predicted field preserves the structure of the ground truth, which is essential for understanding the quality of the downscaled SST field. 
It is computed using three main components: luminance or local means ($\mu_{\hat{y}}$, $\mu_y$), variance (contrast) ($\sigma_{\hat{y}}$, $\sigma_y$) of each field, and covariance (structural similarity) between two fields ($\sigma_{\hat{y}, y}$), which is given as
\begin{linenomath*}
\[
    \mathrm{SSIM} = \frac{(2\mu_{\hat{y}} \mu_y + C_1)(2\sigma_{\hat{y}y} + C_2)}
{(\mu_{\hat{y}}^2 + \mu_y^2 + C_1)(\sigma_{\hat{y}}^2 + \sigma_y^2 + C_2)},
\]
\end{linenomath*}
where constants $C_1 = (0.01)^2$ and $C_2 = (0.03)^2$ are used to stabilize the division in the case of small denominator, assuming unit-normalized SST fields. SSIM ranges from -1 to 1, where values close to 1 indicate high structural similarity.

Another metric used is the $R^2$ score, also known as the coefficient of determination. It measures the proportion of variance in the true values that is captured by the predictions. It is given as
\begin{linenomath*}
\[
R^2 = 1 - \frac{ \sum_{i=1}^{N} (y_i - \hat{y}_i)^2 }{ \sum_{i=1}^{N} (y_i - \bar{y})^2 },
\]
\end{linenomath*}
where $\bar{y}$ the mean of the true values. An $R^2$ value of 1 implies perfect prediction, whereas a value of 0 implies that the model performs no better than predicting the mean.

In addition to these evaluation metrics, we provide an interpretability analysis using Shapley Additive Explanations (SHAP). SHAP enables a detailed understanding of the influence of each input variable on the model predictions. A key advantage of SHAP is its model-agnostic nature, allowing it to be applied to any predictive model regardless of its underlying architecture. The SHAP value for an input is computed as the average marginal contribution of that input across all possible subsets of inputs as
\begin{linenomath*}
\[
    \phi_i = \sum_{S \subseteq F \setminus \{i\}} \frac{|S|! \, (|F| - |S| - 1)!}{|F|!} \left[ f(S \cup \{i\}) - f(S) \right],
\]
\end{linenomath*}
where $\phi_i$ is the SHAP value for an input $i$, $F$ is the set of all inputs, $S$ is a subset of inputs excluding $i$, and $f(S)$ is the model prediction when only the inputs in $S$ are known. This formulation ensures a fair allocation of the contribution among features by averaging over all possible subsets \cite{lundberg2017unified}. SHAP has been used previously to interpret machine learning models in climate research  \cite{descals2023local, silva2022using} .

\section{Results}
In this section, the proposed framework is evaluated using two different test cases, chosen to represent general ocean conditions and extreme marine heatwave conditions, respectively.

First, in Subsection \ref{ModelComp:Generalized}, we assess the robustness, accuracy, and generalization performance of the RCNN model across different training periods, seasons, and spatial regions. All results in this subsection are evaluated on the year 2021, which does not contain a major marine heatwave event. For this generalized case, the standard RCNN is used, as the low-resolution input SST remains within the range of values represented in the training data. Under such conditions, the RCNN is able to downscale SST reliably without requiring additional modifications.

Second, in Subsection \ref{sec:MHWresults}, we focus specifically on the 2011 Western Australian marine heatwave, an extreme event characterized by unusually high SST values that are sparsely represented in the historical training dataset. For this case study, we use the custom loss-assisted RCNN (CL-RCNN) introduced in Section \ref{ImprovementMHW}. This variant is designed to better capture rare and extreme SST anomalies by incorporating synthetic MHW-like data and a custom loss formulation during training.

The use of RCNN versus CL-RCNN is therefore intentional and context-dependent. While CL-RCNN improves performance during extreme events, it requires additional steps such as synthetic data generation and further fine-tuning, which increases computational cost. For routine downscaling applications where no extreme event is indicated in the low-resolution input SST, the standard RCNN is sufficient and computationally more efficient. In contrast, when the input SST suggests the presence of a marine heatwave, the CL-RCNN provides a more appropriate choice. For prospective use, we envision RCNN as the default model for routine downscaling when the coarse-resolution SST remains within the range represented in the training data. CL-RCNN is most appropriate when the coarse-resolution forecast or an external marine-heatwave alert indicates unusually warm SST conditions, or when the application specifically targets upper-tail risk assessment. In operational use, this could be implemented as a simple screening step based on the coarse SST anomaly or percentile threshold: standard RCNN is used for routine conditions, while CL-RCNN is selected when the input field indicates potential MHW-like conditions.

For both test cases, model performance is compared against baseline approaches, including random forest–based interpolation and a standard U-Net. In addition, Subsection \ref{SHAPAnalysis} presents a SHAP-based analysis to examine the contribution of individual input variables, while Subsection \ref{Overfitting} evaluates the generalization behavior and potential overfitting of the proposed RCNN framework.

\subsection{Generalized Case for Non MHW Conditions}
\label{sec:generalCase}
\label{ModelComp:Generalized}
In this subsection, we evaluate the general predictive capabilities of the generalized RCNN (without synthetic training). A detailed analysis focusing on a specific MHW event is provided in Section \ref{sec:MHWresults}. We evaluated the performance of the RCNN model against a random forest-based interpolation method and a standard U-Net. The evaluation focuses on the ability of the models to generalize to out-of-sample conditions. The ROMS data is considered as truth in this work.

To evaluate how the time span of training datasets influences model performance, we tested multiple RCNN models trained on five different time windows and evaluated their predictions against the true SST field.
We begin by examining the sensitivity of RCNN performance to the choice of training period by training multiple models on different temporal windows spanning 2000–2021 and evaluating them on a fixed test year (2021). This analysis was used to assess how the temporal coverage and recency of training data influence generalization performance.

Overall, models trained on longer or more recent datasets exhibited more stable and accurate predictions, while models trained on older periods showed systematic degradation. 
Based on this analysis, we select the 2015–2020 training window for all subsequent experiments in this subsection, as it provides a strong balance between predictive skill and computational efficiency. Unless otherwise stated, all quantitative results reported in Section 3.1, including those in Table~3, correspond to models trained on the 2015–2020 dataset and evaluated on the year 2021.
To isolate architectural effects from those associated with training data selection, we additionally compare RCNN against a standard U-Net trained on the same 2015–2020 dataset. This comparison highlights the benefit of iterative residual correction over a single-pass convolutional model under generalized (non-extreme) conditions.
A detailed quantitative and qualitative analysis of training-period sensitivity is provided in \ref{app:general}.
\begin{table}[htb]
\caption{Performance comparison of models trained on 2015-2020 datasets using RMSE, $R^2$, and SSIM metrics. Metrics are calculated as the mean of daily evaluation scores over the full year for each model for the entire region.}
\label{Tab:ModelPerformance}
\begin{center}
\begin{tabular}{l c c c c c c c c c}
\hline
\textbf{Model} & \multicolumn{3}{c}{\textbf{2021}} \\
\cline{2-4}
 & RMSE & $R^2$ & SSIM \\
\hline
Interpolation &  0.939 & 0.825 & 0.975 \\
U-Net         &  0.515 & 0.911 & 0.983 \\
one-step RCNN & 0.504 & 0.934 & 0.989 \\
RCNN          &  0.389 & 0.955 & 0.998 \\
\hline
\end{tabular}
\end{center}
\end{table}
We also compared the RCNN approach with other models such as interpolation and the base U-Net approach. These models are trained on the 2015-2020 datasets.  The models are evaluated using different metrics, such as RMSE, $R^2$, and SSIM. Table \ref{Tab:ModelPerformance} presents a quantitative comparison of the three SST downscaling approaches evaluated for the year 2021.

For each model configuration, training was repeated using multiple random seeds. The values reported in the main result tables correspond to the best validation-selected model, rather than an average over all seeds. For each seed, the checkpoint with the lowest validation RMSE was retained, and the final model used for the spatial maps, time series, and diagnostic figures was selected based on validation performance. These metrics are calculated as the mean of daily evaluation scores over the full year for each model for the entire region.
Table \ref{Tab:ModelPerformance} clearly shows that interpolation falls short in terms of prediction accuracy and spatial detail compared to the other two approaches. However, the reasonably large values of SSIM (0.975) indicate that interpolation preserves the general spatial structure but lacks the fine-scale details. The U-Net significantly improves on this, but the RCNN model consistently achieves the best results across all metrics. For example, the RMSE for the RCNN model is 0.389 $^\circ$C, while for the U-Net is 0.515 $^\circ$C. This shows a noticeable improvement using the RCNN model. In short, the low RMSE, high $R^2$, and near perfect SSIM confirm the suitability of the RCNN model for high-resolution SST downscaling tasks. Because the main tables report the best validation-selected models, we additionally examined run-to-run variability using multiple random seeds. This analysis confirms that although individual neural-network runs vary with random initialization, the RCNN performance remains consistently better than the U-Net baseline in validation-selected comparisons. Therefore, the improvement reported in Table \ref{Tab:ModelPerformance} is not interpreted as arising from a single random initialization alone, but from the residual correction architecture combined with validation-based model selection.
\begin{figure}[htb]
  \centering
  \includegraphics[width=1\columnwidth]{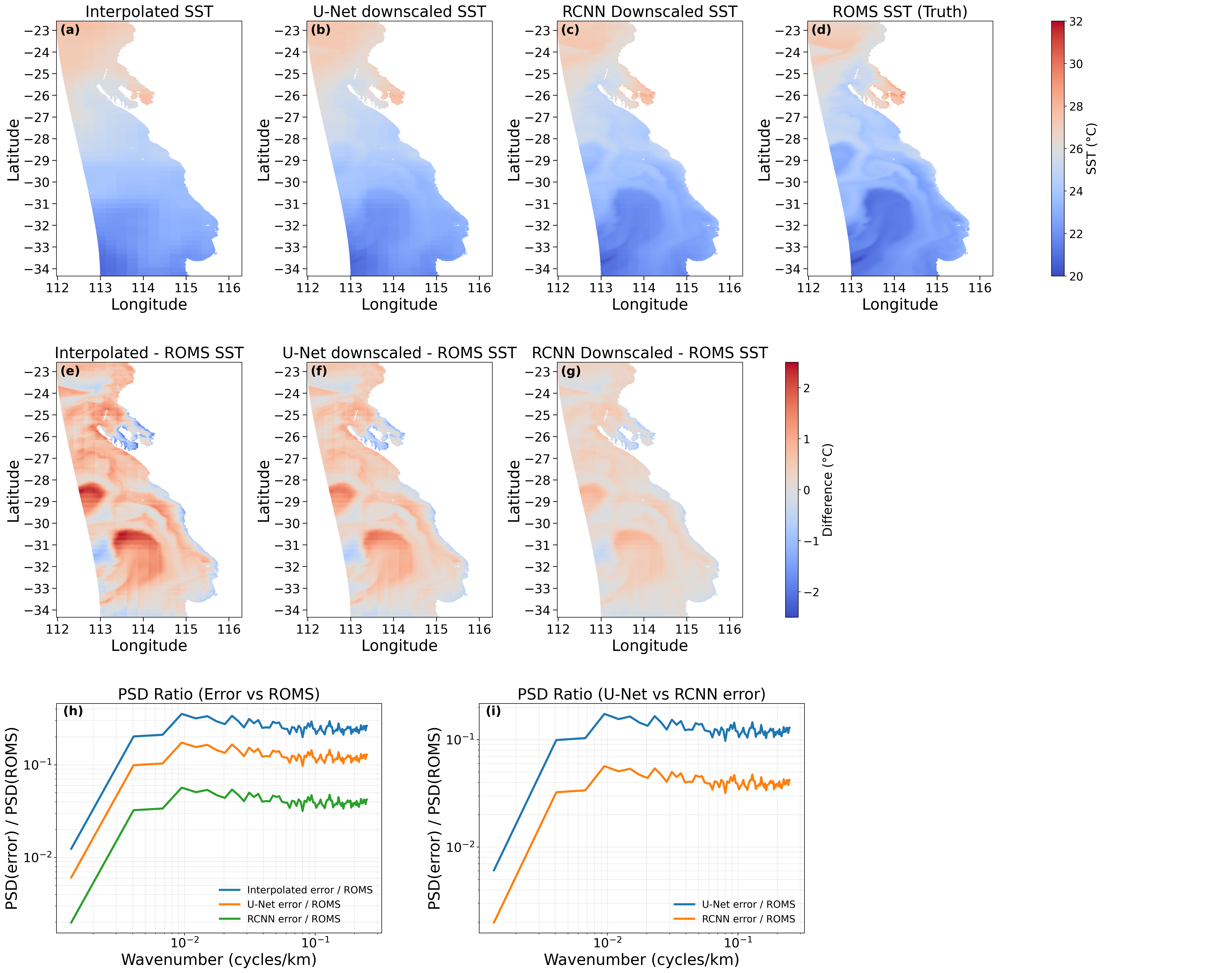}
  \caption{Comparison of sea surface temperature fields on January 31, 2021, along the west coast of Australia. (a) displays the SST interpolated from the coarse-resolution global climate model. (b) presents the SST downscaled using a U-Net model. (c) shows the SST downscaled using an RCNN model. (d) represents the SST from the high-resolution ROMS model, used as the ground truth. (e), (f), and (g) depict the differences between the downscaled SST fields from the Interpolated, U-Net, and RCNN models and the ROMS SST field. (h) and (i) show isotropically averaged power spectral density (PSD) ratios of the reconstruction error relative to the ROMS reference. Lower PSD ratios indicate smaller reconstruction-error energy relative to ROMS at the corresponding spatial scale.
  Note that all models are trained on 2015-2020 dataset. }
  \label{fig:3}
\end{figure}

In addition to the temporally- and spatially-averaged quantitative evaluation, Figure \ref{fig:3} shows the comparison of SST outputs from different downscaling techniques and ROMS ground truth on a single example day, 31.01.2021. Here we focus our attention on the coastal area.

The results of the interpolated baseline model show smooth but oversimplified patterns with loss of finer coastal details, including under-predicting the temperature along the length of the coast. It does not capture the strong gradients in the coastal ocean and has the largest error among all models. The maximum error between interpolated SST and ROMS SST is 2.49 $^\circ C$, as seen in Fig. \ref{fig:3} (e). The SST field obtained using the U-Net model improves resolution. It offers a noticeable improvement in spatial resolution over interpolation, particularly in regions of moderate temperature gradients. However, it still lacks finer-scale realism. The RCNN model further enhances spatial accuracy and provides better structural coherence. It captures fine-scale structures and temperature gradients similar to those observed in the ROMS SST. We note the significant sharpening of the diffuse temperature gradients seen in the interpolated SST. The difference between RCNN downscaled SST and ROMS SST field shows that the RCNN model works well, with the largest difference of 0.897 $^\circ C$.

To further assess the multiscale fidelity of the downscaled fields, Fig. \ref{fig:3} (h–i) shows isotropically averaged power spectral density (PSD) ratios of the error relative to the ROMS reference given by $\mathrm{PSD}(\mathrm{SST}_{model} -\mathrm{SST}_{ROMS})/\mathrm{PSD}(\mathrm{SST}_{ROMS})$.
The interpolated baseline exhibits the largest error energy across all spatial scales, while the U-Net reduces error at intermediate scales but retains elevated error at higher wavenumbers. In contrast, the RCNN consistently achieves the lowest error-to-signal spectral ratios across the resolved wavenumber range, indicating improved recovery of fine-scale coastal variability.
\begin{figure}[htb]
  \centering
  \includegraphics[width=1\columnwidth]{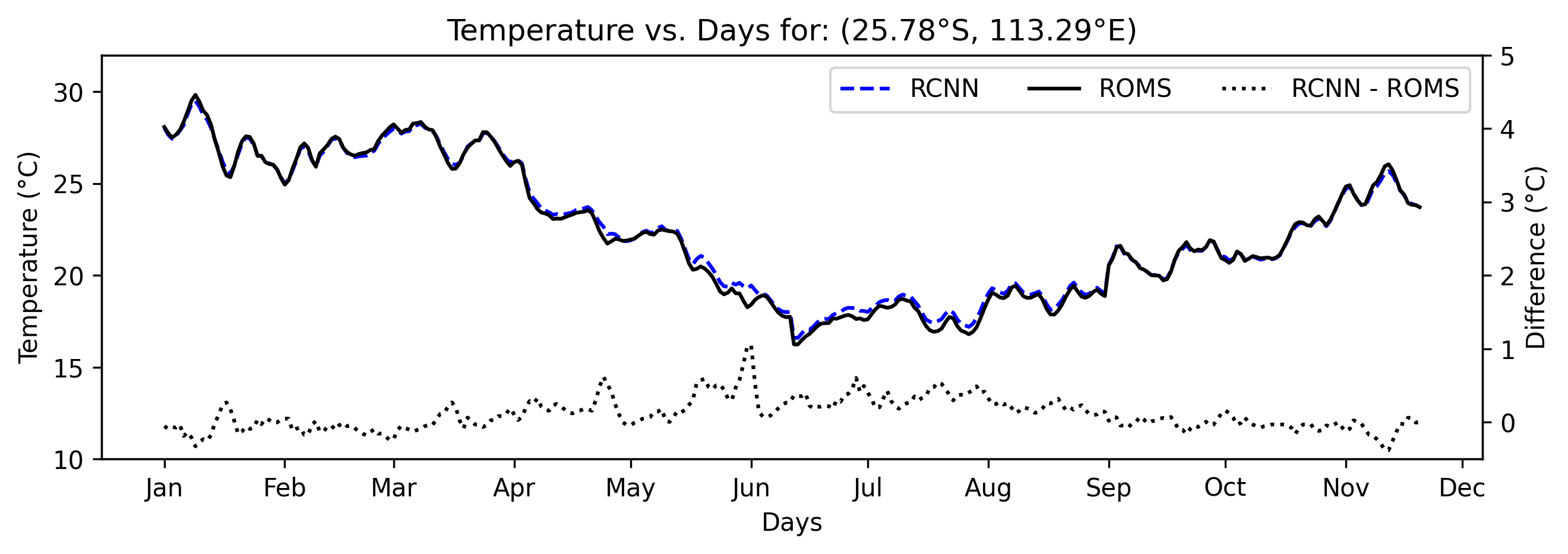}
  \caption{Comparison between the RCNN model trained on 2015-2020 dataset and ROMS based on daily sea surface temperature time series at the coastal location (25.78$^\circ$S, 113.29$^\circ$E) for the year 2021. The difference between the RCNN SST and ROMS SST is shown as a dotted black line on the right-hand y-axis.}
  \label{fig:4}
\end{figure}

To understand the quality of the SST downscaled by RCNN, we plotted the daily SST time series at the coastal  Shark Bay location (25.78$^\circ$S, 113.29$^\circ$E) for the entire year of 2021 in Fig. \ref{fig:4}. 

The plot compares the RCNN downscaled SST (dashed blue line) with the ROMS SST (solid black line). The RCNN SST follows the ROMS SST very closely with minimum error throughout the year. The difference remains small and is generally less than $\pm 1^\circ$C, suggesting high accuracy. Seasonal patterns are also well captured by the RCNN model, where warmer SST occurs in the first 100 days, a cooling period in the middle, and then gradual warming. Additionally, it is evident that the RCNN can preserve the short-term fluctuations relatively well and does not smooth the results. 
\subsection{Case Study: 2011 Marine Heatwave Event}
\label{sec:MHWresults}
The 2011 marine heat wave off the west coast of Australia was a significant ocean warming event characterized by abnormally high SSTs that occurred in early 2011. This event was primarily driven by a stronger than usual Leeuwin Current, which transported warm tropical waters southward along the coast \cite{pearce2013rise}. Positive SST anomalies of around 3 $^\circ$C were observed in some regions; significantly affecting marine ecosystems and several commercial fisheries. The heatwave led to widespread coral bleaching, seagrass and kelp loss, changes in the distributions of marine species, and declines in commercially important fisheries, such as abalone and scallops \cite{pearce2011marine}. Therefore, given the ecological and economical impact of the 2011 MHW event, it serves as a critical test case for evaluating the performance of statistical downscaling methods for SST forecasting.
Three different methods namely, random forest-based interpolation, U-Nets, and CL-RCNN described in Subsection \ref{subsec:RCNN} were used to downscale the SST of 2011 to the ROMS grid. 

To optimize model performance and minimize computational cost, the training dataset is selected dynamically by successively increasing the numbers of years included in the dataset. While larger training datasets generally improved validation accuracy, comparable performance was obtained using a reduced training window with substantially lower computational cost. A detailed analysis is provided in \ref{app:mhw}.
Based on this analysis, all CL-RCNN results reported in this subsection correspond to models trained on the 2005–2010 dataset.

Note that for this marine heatwave case study, the custom loss-assisted RCNN (CL-RCNN) model is used as described in Section \ref{ImprovementMHW}. This model is specifically designed to improve the prediction of extreme SST anomalies associated with MHW events. In order to generate the synthetic data, we have used the methodology described in Subsection \ref{ImprovementMHW}. The reference ROMS field used for injecting the modulated Gaussian features is gathered from the year 2010. 
\begin{figure}[htb]
  \centering
  \includegraphics[width=1\columnwidth]{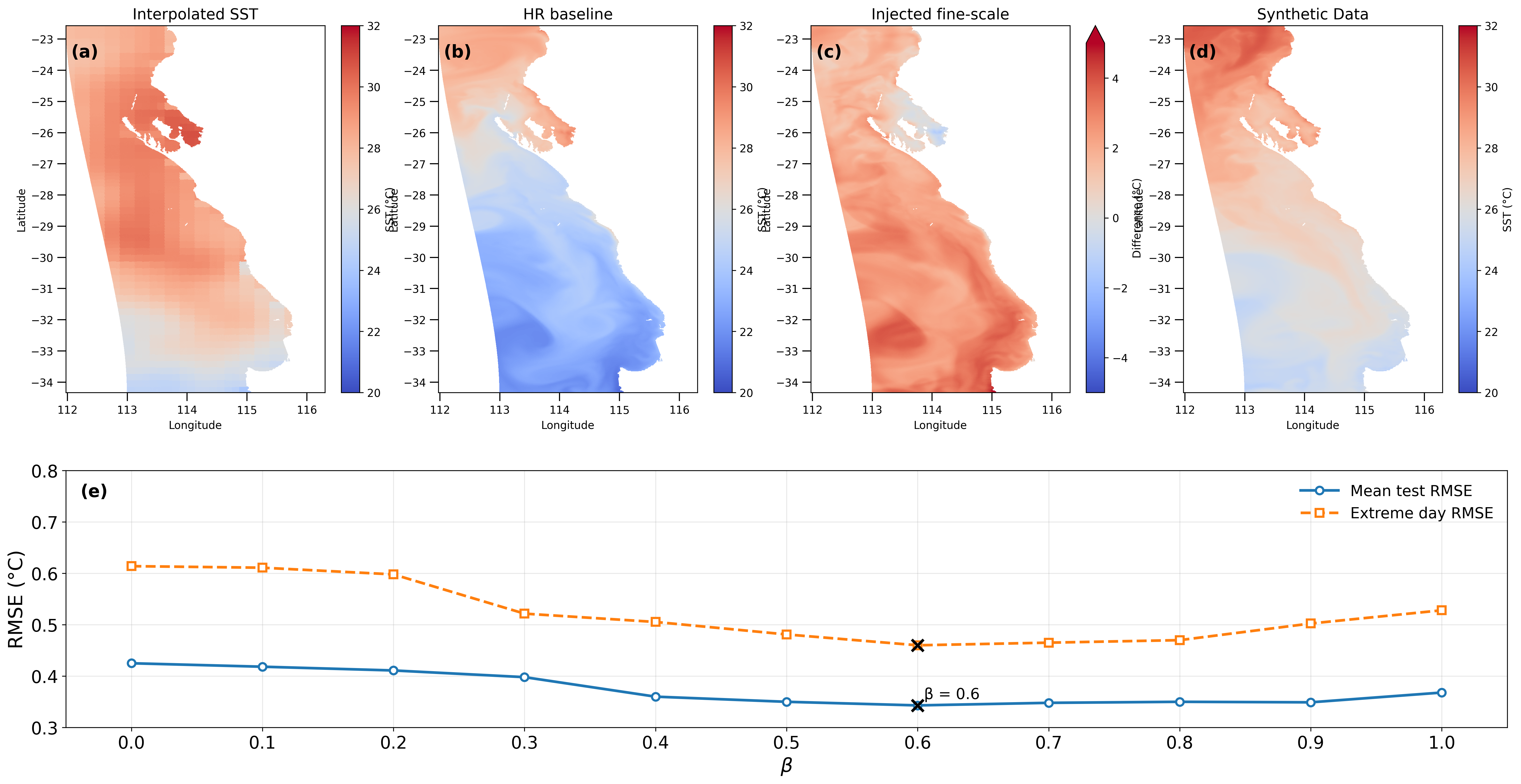}
  \caption{Construction of synthetic sea surface temperature (SST) fields and sensitivity of the custom loss formulation to the weighting parameter $\beta$. Panels (a–d) show, for a representative day in February 2010, (a) the interpolated coarse-resolution SST, (b) the high-resolution ROMS baseline field, (c) the injected fine-scale perturbations, and (d) the resulting synthetic SST field used for training the custom loss-assisted RCNN. Panel (e) shows the dependence of RMSE on $\beta$ for both the mean test period and the worst marine heatwave day. Black crosses denote the minimum RMSE attained for each curve, indicating optimal performance for intermediate $\beta$ values.}
  \label{fig:5}
\end{figure}
Figure \ref{fig:5} illustrates both the synthetic SST data generation process and the sensitivity of model performance to the custom loss weighting parameter $\beta$. Panels (a–d) show the interpolated SST, the high-resolution ROMS baseline field, the injected fine-scale perturbations, and the resulting synthetic SST field, respectively, for a representative day in February 2010.

The synthetic SST field (panel d) preserves the large-scale thermal structure of the ROMS baseline while introducing spatially coherent fine-scale variability representative of marine heatwave conditions. Importantly, the synthetic fields expand the effective SST range relative to the baseline ROMS data, enhancing the representation of rare high temperature states associated with MHWs. For the example shown, the baseline ROMS SST spans 20.86-28.88 $^\circ$C, whereas the synthetic field reaches values up to 31.05 $^\circ$C, comparable to the MHW-affected interpolated SST from 2011 (23.93-30.85 $^\circ$C), but without introducing unphysical extremes. This controlled range expansion enables the model to encounter high-SST conditions during training that are underrepresented in the historical ROMS dataset.
To determine the optimal weighting of synthetic data during training, panel (e) demonstrates the model's RMSE as a function of $\beta$ for both the mean test period (January–March 2011) and the most extreme MHW day (15 February 2011). A clear U-shaped dependence emerges for both metrics, with minimum RMSE of 0.343 $^\circ$C (mean) and 0.460 $^\circ$C (extreme day) attained at $\beta \approx 0.6$. At $\beta = 0$, corresponding to the standard RCNN without synthetic augmentation, the model exhibits relatively higher RMSE values (0.425 °C for the mean test period and 0.614 °C for the extreme day; see Table \ref{Tab:ModelPerformance2011}), highlighting the quantitative benefit of incorporating synthetic data. Conversely, excessive weighting ($\beta > 0.8$) degrades performance due to overrepresentation of synthetic spatial correlations that differ from real observations.

The broad optimum observed for $\beta \in [0.3, 0.8]$ indicates robustness to exact hyperparameter selection, with differences in RMSE across this range being minimal. Based on this analysis, we adopt $\beta = 0.6$ for all subsequent CL-RCNN results, achieving a 19.4\% RMSE reduction for standard conditions and 25.1\% for the extreme MHW day relative to the standard RCNN approach. This validates the synthetic data augmentation strategy for improving model performance on both typical and rare extreme events. To assess whether the CL-RCNN improvement depends on a single synthetic-data configuration, we provide a sensitivity analysis over the Gaussian smoothing scale $\sigma_G$ and synthetic anomaly injection strength $\gamma$ in \ref{app:synthetic_sensitivity}.

\begin{table}[htb]
\caption{Performance comparison of models for January--March 2011 and the extreme day of 15 February 2011. The Pixel Extreme RMSE is computed over pixels where ROMS $\mathrm{SST} > 28^\circ \mathrm{C}$, while Pixel Non-Extreme RMSE is computed over pixels where ROMS $\mathrm{SST}\leq 28^\circ$C. Note that all models are trained on the 2005--2010 dataset.}
\label{Tab:ModelPerformance2011}
\scriptsize
\begin{center}
\begin{tabular}{l c c c c c c}
\hline
\textbf{Model} & \multicolumn{3}{c}{\textbf{Jan--Mar 2011}} & \textbf{Pixel} & \textbf{Pixel} & \textbf{Extreme-day} \\
\cline{2-4}
 & RMSE & $R^2$ & SSIM & \textbf{Extreme RMSE} & \textbf{Non-Extreme RMSE} & \textbf{RMSE} \\
\hline
Interpolation & 1.005 & 0.804 & 0.968 & 1.058 & 0.987 & 1.032 \\
RCNN          & 0.425 & 0.897 & 0.979 & 0.588 & 0.418 & 0.614 \\
CL-RCNN       & 0.343 & 0.950 & 0.992 & 0.410 & 0.320 & 0.460 \\
\hline
\end{tabular}
\end{center}
\end{table}

The performance Tab. \ref{Tab:ModelPerformance2011} shows the comparison between three models, namely the interpolation model, RCNN and CL-RCNN based on $R^2$, RMSE and SSIM index for the period of January 2011 to March 2011. The table shows that CL-RCNN outperforms all models with the lowest RMSE (0.343 $^\circ$C), the highest $R^2$ (0.950), and the best SSIM (0.992), showing its strength in capturing both general patterns and extreme SST events.
The CL-RCNN achieves an RMSE of 0.343$^\circ$C, which is approximately 20\% lower than the RCNN's RMSE of 0.425$^\circ$C, indicating a notable reduction in overall prediction error. It also gives a higher $R^2$ value of 0.950 compared to 0.897, reflecting an improved ability to explain the variance in the SST data. The SSIM score for CL-RCNN is 0.992, surpassing the RCNN’s 0.979, demonstrating superior preservation of spatial structure, particularly for complex coastal features.

The difference in model performance becomes even more evident on the extreme day of February 15, 2011. On this day, the CL-RCNN records an RMSE of 0.460 $^\circ$C, improving upon the RCNN’s 0.614 $^\circ$C. Furthermore, $R^2$ and SSIM values remain higher, 0.941 and 0.989 for CL-RCNN, compared to 0.876 and 0.970 for RCNN. These results highlight the advantage of the CL-RCNN, which enhances the model’s ability to capture sharp gradients and temperature extremes that the standard RCNN tends to underestimate. 

To directly evaluate whether CL-RCNN maintains performance over the standard temperature range, we additionally computed RMSE over non-extreme pixels, defined as locations where ROMS SST $\leq 28^\circ$C. The non-extreme RMSE decreases from 0.418$^\circ$C for RCNN to 0.320$^\circ$C for CL-RCNN. At the same time, the upper-tail Pixel Extreme RMSE decreases from 0.588$^\circ$C to 0.410$^\circ$C. These results indicate that the synthetic-data retraining improves the representation of high-temperature values without degrading performance over the broader non-extreme SST range.

\begin{figure}[htb]
  \centering
  \includegraphics[width=1\columnwidth]{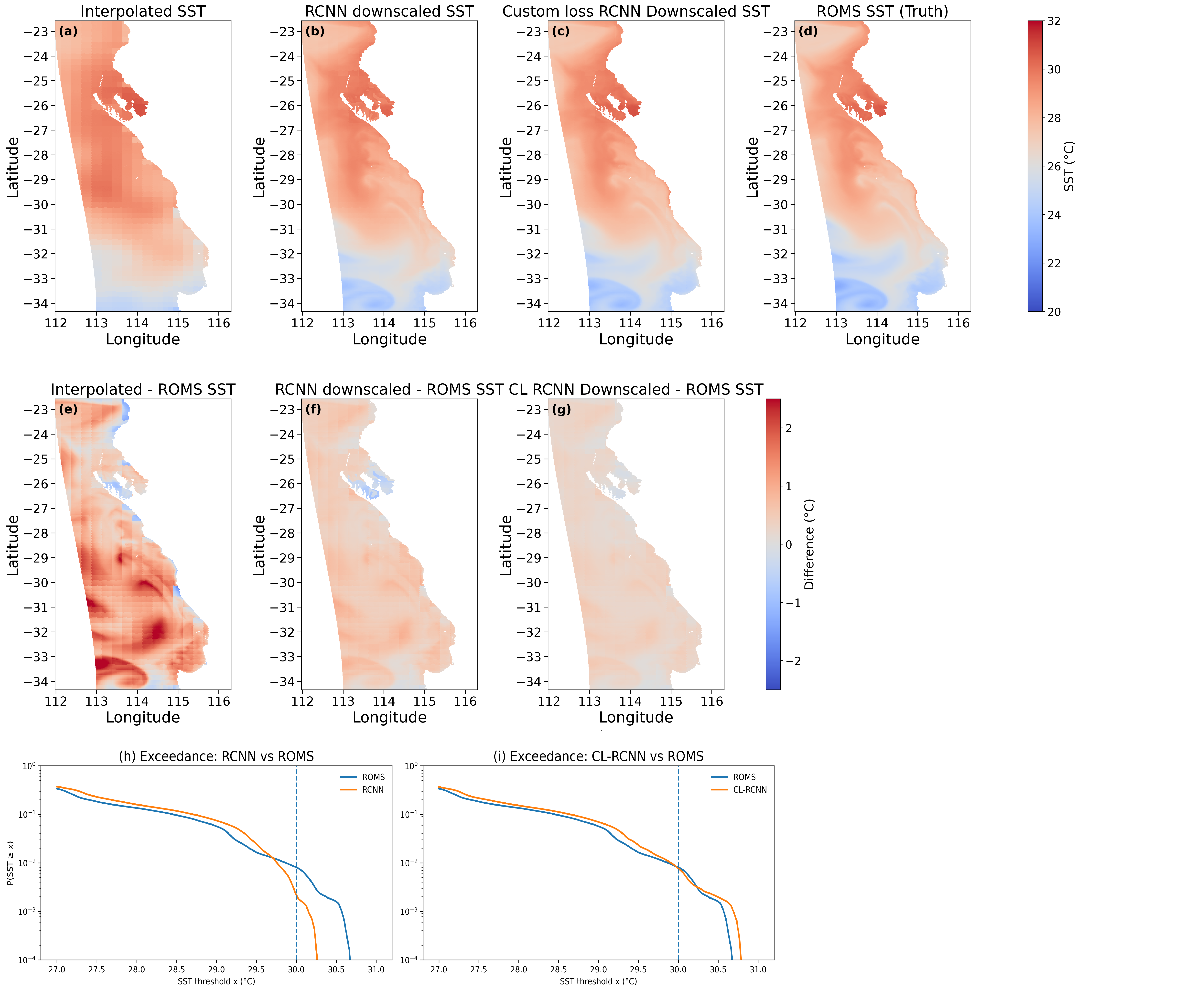}
  \caption{Comparison of sea surface temperature fields on February 15, 2011, along the west coast of Australia. (a) displays the SST interpolated from the coarse-resolution global climate model. (b) presents the SST downscaled using a RCNN model. (c) shows the SST downscaled using an custom loss-assisted RCNN model. (d) represents the SST from the high-resolution ROMS model, used as the ground truth. (e), (f), and (g) depict the differences between the downscaled SST fields from the Interpolated, RCNN, and CL-RCNN models and the ROMS SST field. Note that all models are trained on 2005-2010 dataset.}
  \label{fig:6}
\end{figure}

For spatial comparison, Fig. \ref{fig:6} shows the comparison between the SST fields on 15 February 2011 obtained using the RCNN and CL-RCNN models. 
The SST field obtained using the RCNN model plotted in subplot (b) shows that the model is able to recover the fine-scale structures quite well. Subplots (e) and (f) demonstrate that errors are significantly reduced for the RCNN model compared to interpolation. However, in panel (f), corresponding to the standard RCNN downscaling, a prominent region of negative differences (blue patch) is observed near the coast around the Shark Bay area ($\sim 26^\circ$S). This indicates that the model underestimates SST in this area compared to the ROMS truth.  Specifically, the RCNN model is seen to struggle to accurately predict temperature values above $30^\circ$C. The problem is that temperatures above $30^\circ$C only occur in small regions or for short periods in the training dataset. Because they are rare, the model sees very few examples of these high values during training. As a result, it is biased towards predicting the more common moderate temperatures and fails to generalize to extremes.

The CL-RCNN model overcomes this problem. Subplot (g) shows the difference between the ROMS SST and the custom loss-assisted RCNN downscaled SST. We can see that the regions where the RCNN model previously underestimated the SST (blue patches) are smaller and less intense in subplot (g). This improvement indicates that the CL-RCNN model predicts higher temperature values more accurately, especially in regions where extreme SST values are observed.

This behaviour is further confirmed by the exceedance probability analysis shown in panels (h) and (i). Here, the x-axis represents a temperature threshold $x$, and the y-axis shows the fraction of grid points with SST values greater than or equal to this threshold. Panel (h) demonstrates that the exceedance probability for the standard RCNN drops off rapidly beyond approximately 30 C, highlighting its tendency to under-represent rare warm extremes. In contrast, panel (i) shows that the CL-RCNN follows the ROMS exceedance curve much more closely across the warm tail of the distribution, confirming that the custom loss formulation improves the representation of extreme SST values that are not evident from spatial maps alone.

\begin{figure}[htb]
  \centering
  \includegraphics[width=1\columnwidth]{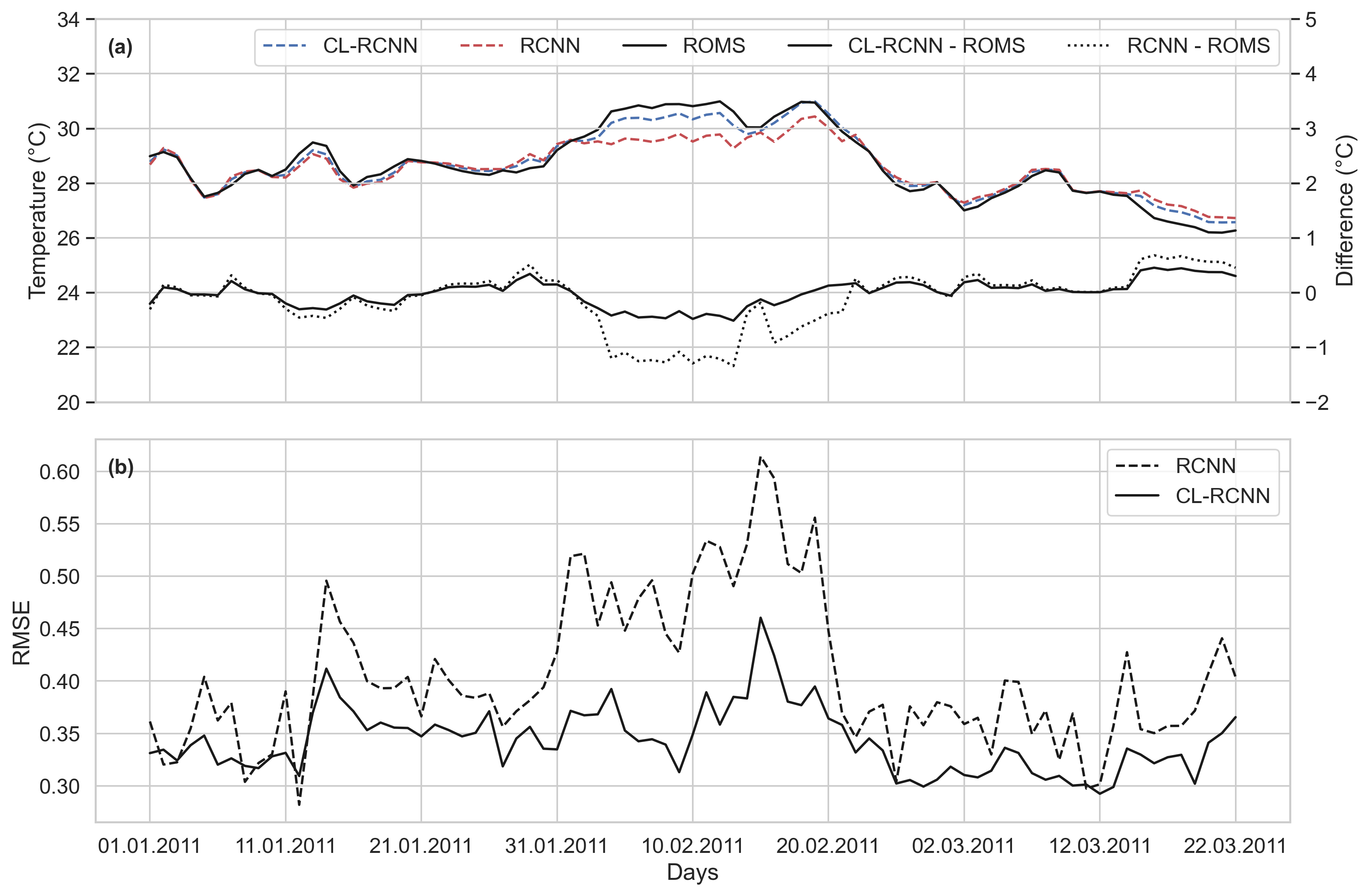}
  \caption{(a) Comparison between the RCNN model trained on 2005-2010 dataset and ROMS based on daily sea surface temperature time series at the coastal location (25.78$^\circ$S, 113.29$^\circ$E) from January to March 2011. The difference between the RCNN SST and ROMS SST is shown as a dotted black line on the right-hand y-axis. The difference between the CL-RCNN SST and ROMS SST is shown as a solid black line on the right-hand y-axis. (b) Performance comparison between RCNN and CL-RCNN from January to March 2011 for the daily SST field based on the RMSE between the models outputs and ROMS.}
  \label{fig:7}
\end{figure}

Furthermore, to visualize how models behave over time and clearly show the benefit of the CL-RCNN method, the time series (Jan-March 2011) of SST at the coastal location (25.78$^\circ$S, 113.29$^\circ$E) is plotted in Fig. \ref{fig:7} (a).
\begin{figure}[htb]
  \centering
  \includegraphics[width=1\columnwidth]{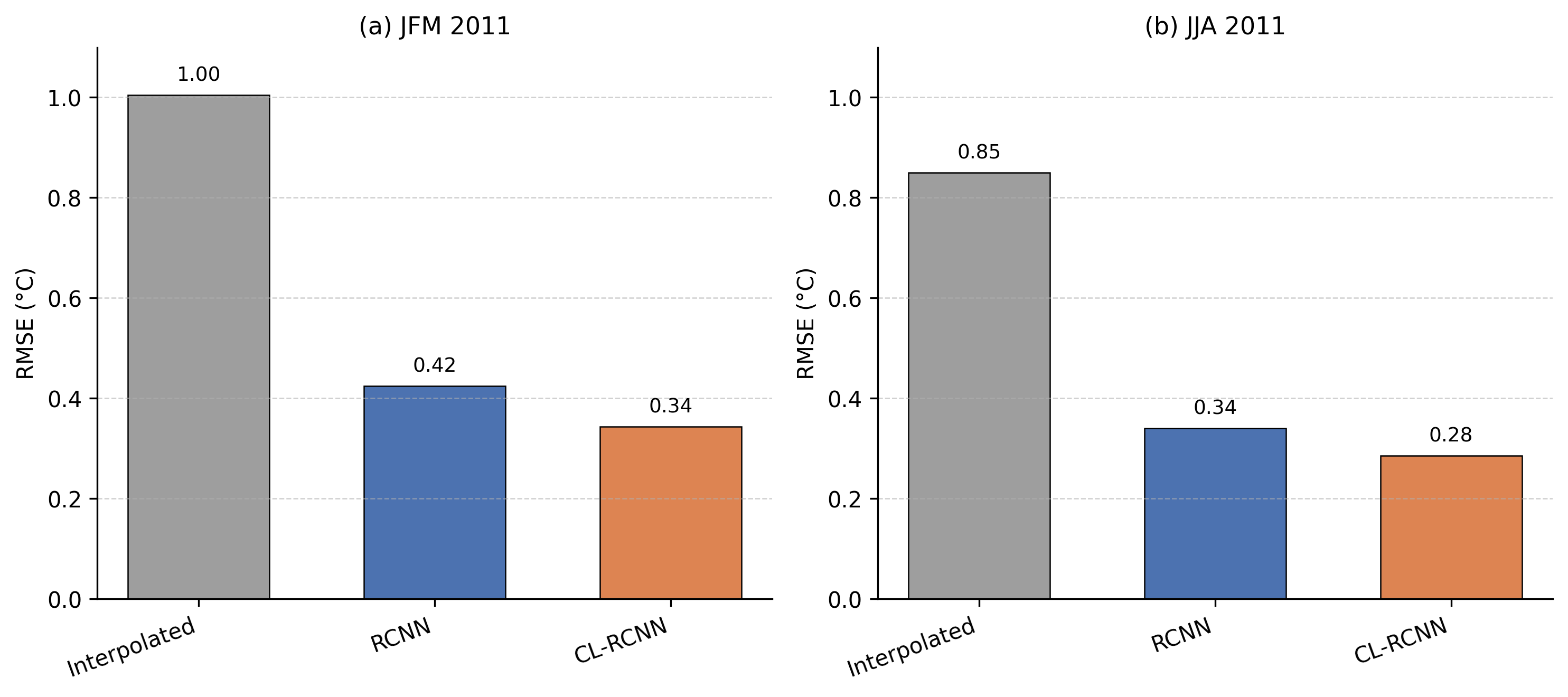}
  \caption{Seasonal evaluation of SST reconstruction performance during 2011. Spatially averaged root-mean-square error (RMSE) relative to the high-resolution ROMS reference is shown for (a) austral summer (JFM) and (b) austral winter (JJA).}
  \label{fig:8}
\end{figure}
Throughout the time period, the CL-RCNN prediction (blue dashed line) consistently tracks the ROMS SST more closely than the standard RCNN (red dashed line), especially during the peaks of high-temperature events. The RCNN model underestimates the high temperature and shows large differences, while the CL-RCNN model shows smaller differences. This shows that the methodology based on the custom loss function significantly improves the performance of the model. Figure \ref{fig:7} (b) also shows the evolution of RMSE over time. It is evident that the RMSE values are consistently higher for the RCNN model compared to the CL-RCNN model, indicating improved performance with the use of the custom loss function.

\begin{figure}[htb]
  \centering
  \includegraphics[width=1\columnwidth]{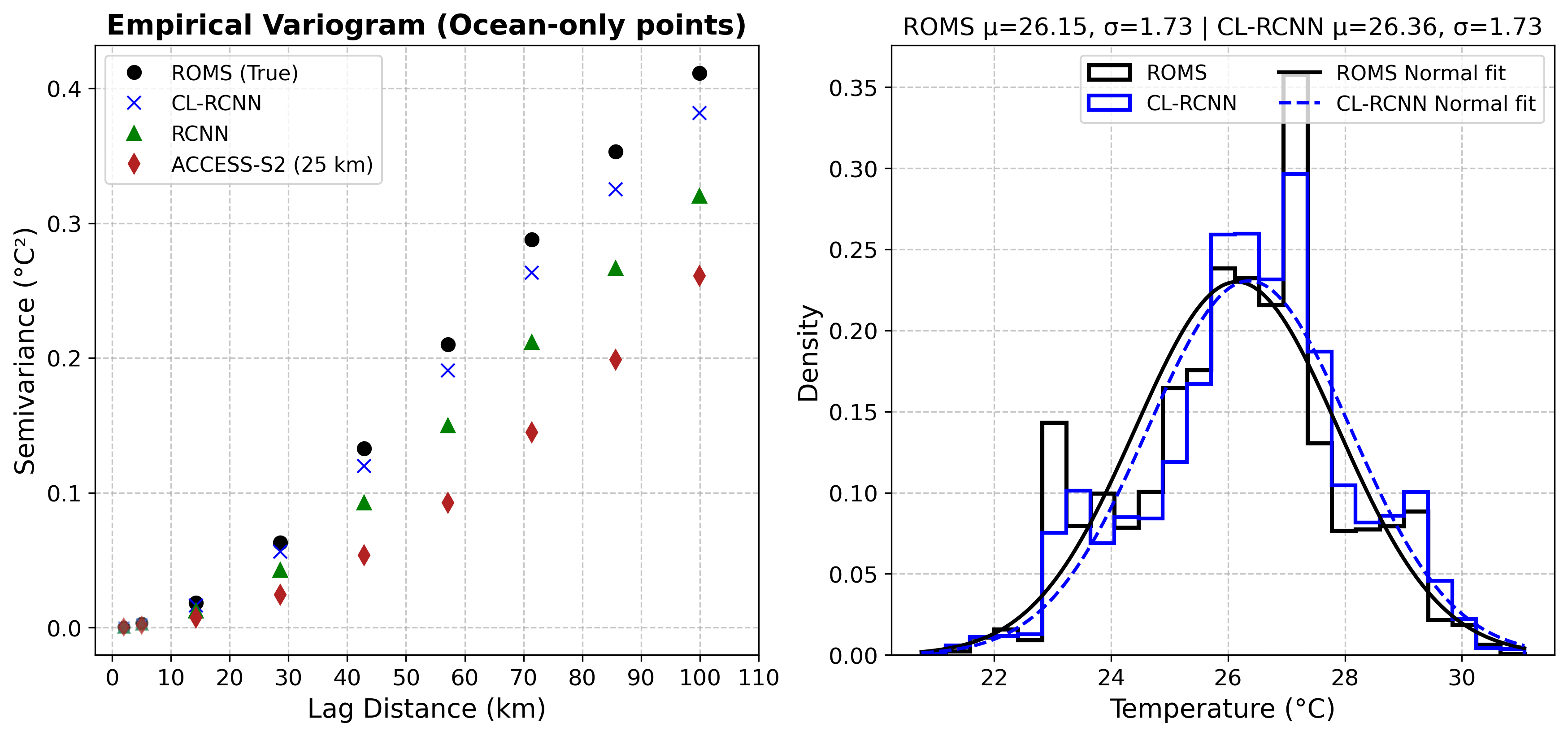}
  \caption{Comparison of the spatial structure and temperature distribution between the true ROMS field and the CL-RCNN prediction on February 15, 2011. (a) is the empirical variograms showing the semivariance as a function of lag distance for different models. (b) is the histograms and probability density functions of temperature values. Note that the CL-RCNN model is trained on 2005-2010 dataset}
  \label{fig:9}
\end{figure}

While such time-local diagnostics are useful for illustrating individual events, they do not by themselves indicate whether the observed variability reflects a systematic seasonal dependence of the method. 

Therefore, to assess potential seasonal effects more comprehensively, we extend the evaluation to seasonal analysis for austral summer and winter of 2011. Figure \ref{fig:8} shows the spatially averaged RMSE for January–March (JFM) and June–August (JJA) 2011. Consistent with the time-series behavior in Figure \ref{fig:7}, the RMSE values are higher during JFM than during JJA for all methods, reflecting the higher variability of SST, sharper coastal gradients and enhanced mesoscale activity characteristic of the summer season. Despite this increased summer complexity, both RCNN and CL-RCNN substantially outperform simple interpolation in both seasons, with CL-RCNN achieving the lowest RMSE in JFM and JJA. We noticed that CL-RCNN reduces seasonal RMSE by approximately 65\% in JFM and 66\% in JJA compared to the interpolation model, indicating that the majority of interpolation errors are removed in both summer and winter. 
Together, Figures \ref{fig:7} and \ref{fig:8} demonstrate the performance gains of CL-RCNN across seasons.

To further demonstrate the capacity of the CL-RCNN model to predict the finescale variability, Fig. \ref{fig:9} plots the variogram and distribution of the temperature field on February 15, 2011.  Figure \ref{fig:9} (a) compares the empirical variograms of the ACCESS-S2 SST field (25 Km), CL-RCNN, RCNN and true field ROMS. Variograms quantify how spatial variability changes with distance \cite{oliver2014tutorial}. A low semivariance at short distances indicates spatial smoothness, while higher values at larger distances reflect greater differences between the values.
The black circles represent the true ROMS field variogram, while the red diamonds represent the variogram of the coarser global SST field, which shows noticeably higher semivariance. Also, semivariance values for the global field are absent for shorter lag distances (0-25 km), as it does not provide any information at such fine spatial scales. Unlike the ROMS and CL-RCNN fields, the global field exhibits smoother, larger-scale variability, and under-represents submesoscale spatial structures, reinforcing the importance of downscaling methods for coastal analysis. 

The blue crosses represent the CL-predicted field variogram. CL-RCNN and ROMS variograms match closely across all lag distances, indicating the satisfactory performance of the CL-RCNN model. It also demonstrates that the model can mimic spatial structures and variability at shorter distances quite well. Minor deviations at larger distances suggest slightly increased variability in the predicted field, but the overall match is excellent. 

Furthermore, Fig. \ref{fig:9} (b) compares the probability density functions (PDF) and the histograms of the temperature values. Both distributions align closely, capturing the primary mode around $27-28^\circ$C. The mean and standard deviation values are also very close. The mean and standard deviations of the ROMS are $27.44^\circ$C and $1.57^\circ$C, while the mean and standard deviation of the CL-RCNN temperature field are $27.61^\circ$C and $1.53^\circ$C. This indicates that the CL-RCNN model preserves not only the spatial variability but also the overall temperature distribution of the true field.

\begin{figure}[htb]
  \centering
  \includegraphics[width=1\columnwidth]{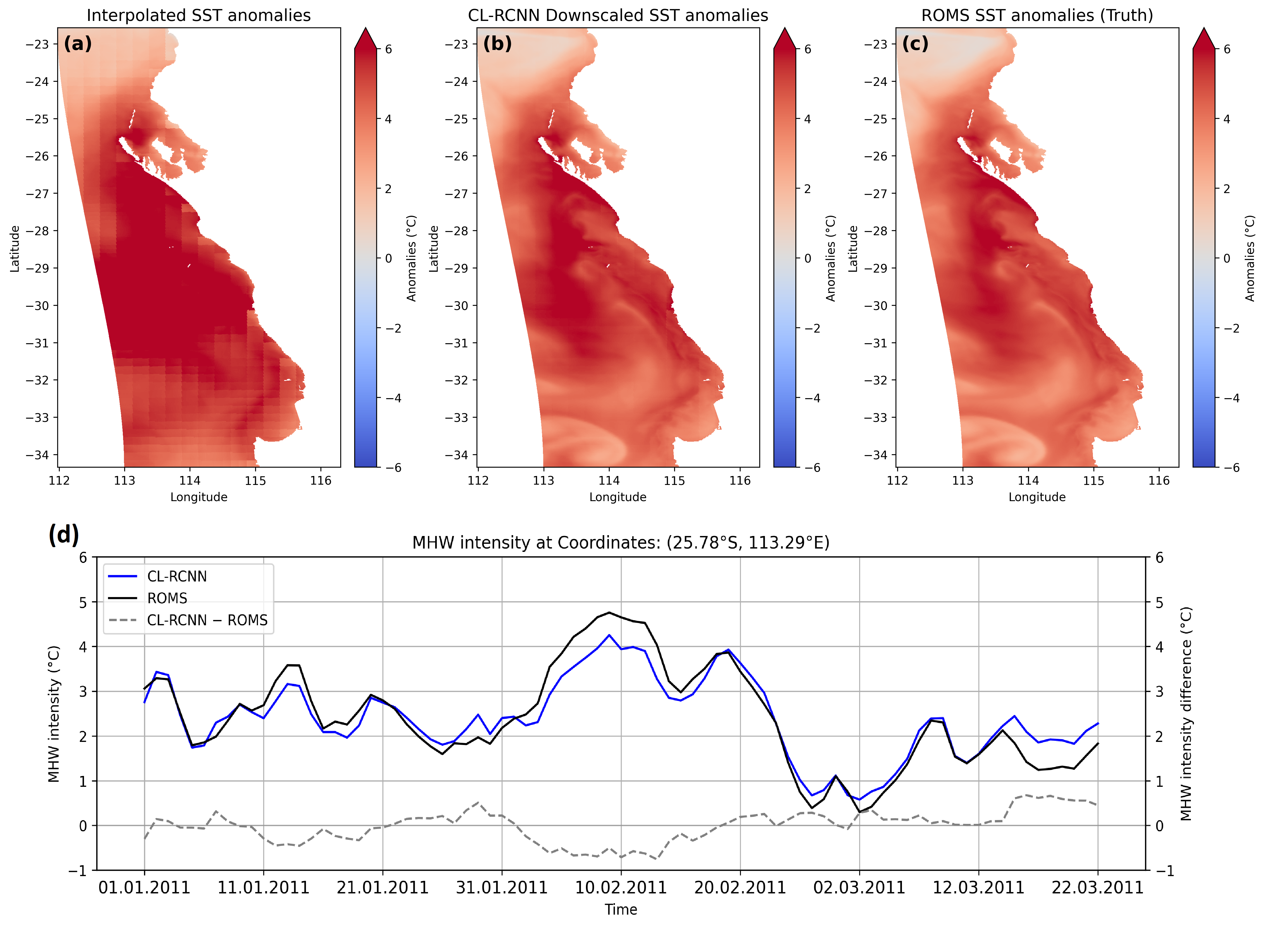}
  \caption{Subplots (a), (b), and (c) show SST anomalies on February 15, 2011, computed as the difference between the SST fields downscaled using different ML models and the climatological mean, highlighting the marine heatwave (MHW) conditions along the Western Australian coast. Subplot (d) shows the variation of SST anomaly from January to March 2011 at a position within Shark Bay (25.78$^\circ$S, 113.29$^\circ$E). The right y-axis shows the difference between the CL-RCNN and ROMS MHW intensity indices, providing a direct assessment of model skill in reproducing the magnitude and temporal evolution of the marine heatwave. Note that all models are trained on 2005-2010 dataset}
  \label{fig:10}
\end{figure}
To highlight the marine heat wave conditions along the Western Australian coast, Figure \ref{fig:10} illustrates the SST anomalies, calculated as the difference between the daily SST predicted by different models and the climatological mean. To compute the climatological mean, we calculated the average of daily SST fields over the period from 2000 to 2011. This multiyear average provides a baseline representation of typical SST conditions, against which anomalies and marine heatwave events can be identified. Subplots (a), (b) and (c) show the SST anomaly fields on February 15, 2011, while subplot (d) shows the variation of SST anomaly from January to March 2011 at coordinates (25.78$^\circ$S, 113.29$^\circ$E). Subplot (a) shows the SST anomalies derived from the interpolated SST field. We can see that high anomaly values are present, but small-scale structures and coastal gradients are poorly resolved. Compared to interpolation, the CL-RCNN prediction captures much finer spatial details and sharper gradients, as illustrated in Subplot (b). Additionally, subplot (d) shows the CL-RCNN closely tracks the temporal evolution of the true MHW anomaly. This indicates the model's ability not only to predict spatial structures but also to reproduce the temporal dynamics of MHW events at specific locations. In short, the CL-RCNN model enhances the ability to capture small-scale temperature structures critical for detecting and predicting the impact of marine heatwaves in coastal regions.

\subsection{Variable importance and spatial SHAP analysis}
\label{SHAPAnalysis}
\begin{figure}[htb]
  \centering
  \includegraphics[width=1\columnwidth]{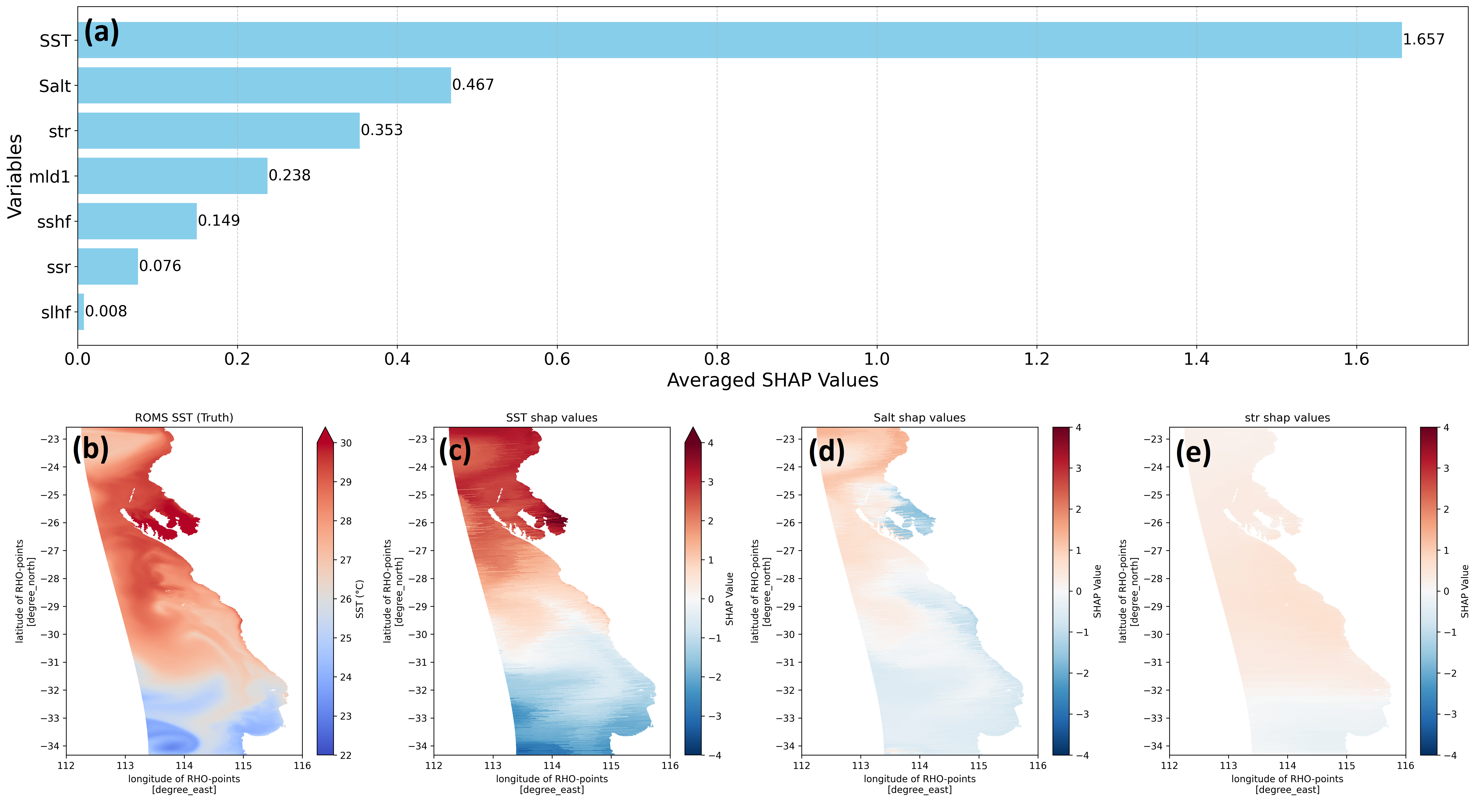}
  \caption{Feature importance and spatial SHAP analysis for SST prediction on February 15, 2011. Subplot (a) shows the mean absolute SHAP values for each input feature, computed by averaging the absolute SHAP values across the entire domain for the day. The four subplots (b), (c), (d), and (e) at the bottom provide a ROMS SST field and the SHAP values for the three most important features, namely Global SST, salinity (salt), and net surface thermal (longwave) radiation (str).}
  \label{fig:11}
\end{figure}
SHapley Additive exPlanations analysis provides an interpretable way to understand the inner workings of a machine learning model. SHAP quantifies the importance of each input variable in contributing to the model’s prediction \cite{lundberg2017unified}. Furthermore, spatial SHAP plots reveal where each variable has more or less influence, supporting regional process understanding. Therefore, in this work, SHAP analysis is performed to make the developed black-box ML models more transparent. 

Figure \ref{fig:11} illustrates the global and spatial contribution of the input variables to the SST prediction made by the CL-RCNN model using SHAP analysis. 
The bar graph (subplot \ref{fig:11}(a)) shows the average absolute SHAP values for all seven input variables, computed by taking the mean of the absolute SHAP values across all spatial grid points on February 15, 2011. This gives a global measure of the importance of the variables for that specific day. This spatial average provides a useful summary of the variables that had the greatest influence on the prediction of the model on that day. Although SHAP values are inherently local, spatial averaging helps to identify dominant drivers regionally.

The SST variable has the highest mean SHAP value (1.657), confirming its dominant role in driving the model’s predictions. Naturally, the SST variable contributes most to the prediction of the target variable since the task is to predict the SST. Salinity (salt) and net surface thermal (longwave) radiation (str) are the next most important variables, with a noticeable influence on the model's performance. Salinity modulates how heat is stored or released, especially in the nearshore and shelf regions. Likewise, str represents the balance of longwave radiation emitted and absorbed at the ocean surface. Since longwave radiation is governed by SST through blackbody emission, str closely reflects the thermal state of the upper ocean. 
Therefore, the relatively high SHAP values of salinity and net surface thermal radiation reflect their role in SST predictions, which is both physically and spatially consistent. 

The spatial distribution of SHAP values is also shown for the three most important variables, together with the reference SST field from ROMS in Fig. \ref{fig:11}. SHAP values are either positive or negative. These values at each point explain how much a particular variable pushes the prediction of the model up or down at that specific point on the grid. For example, if the input variable is high and the SHAP value is positive, then that high variable value pushes the prediction higher. In short, the SHAP value for any variable tells us how much the input variable at that location shifts the prediction away from the global mean. Subplot \ref{fig:11}(c) visualizes the local contribution of the SST variable to the model output in the spatial domain. 
Strong positive SHAP values (red) indicate regions where the SST input increases the model’s prediction relative to the global mean SST prediction, suggesting a local warming influence. Conversely, negative SHAP values (blue) represent regions where the SST input lowers the predicted SST compared to the global mean, indicating a local cooling effect.
In particular, a higher influence is observed in northern coastal areas, suggesting that strong SST gradients are driving predictions in that region. 
Subplot \ref{fig:11}(d) shows the contribution of salinity to the SST prediction. In shallow coastal environments such as Shark Bay, salinity variability is closely linked to evaporation, restricted exchange with the open ocean, and density-driven circulation. These processes affect local heat retention and dispersion, leading to fine-scale SST variability that is explicitly resolved in the high-resolution ROMS simulations but absent from coarse global models. The elevated  importance of salinity therefore indicates that the CL-RCNN is learning to associate salinity-driven coastal and shelf dynamics with localized SST structure. Similarly, subplot \ref{fig:11}(e) shows the SHAP values for str. Although str contributes less to the SST prediction than SST or salinity variables, its SHAP values display coherent spatial patterns that correspond to variations in SST.

In general, by combining global feature ranking with spatial attribution, we can say that SHAP analysis provides a transparent and physically meaningful interpretation of the model's decision process.

\subsection{Generalization and Overfitting}
\label{Overfitting}
\begin{figure}[htb]
  \centering
  \includegraphics[width=0.7\columnwidth]{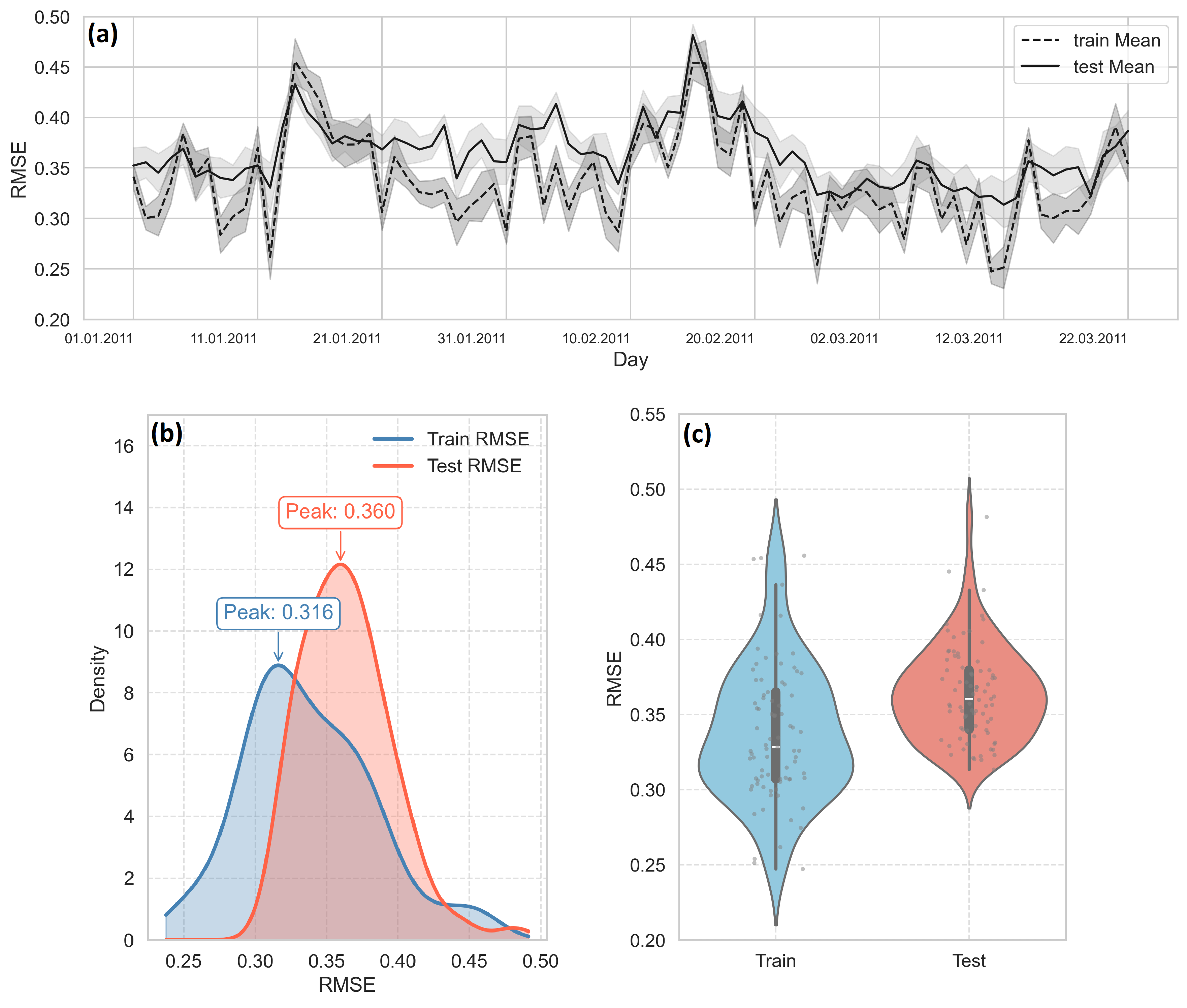}
  \caption{Overfitting and generalization assessment of the CL-RCNN model across 100 random seed runs.
(a) Daily RMSE averaged across 100 different runs for the training period (January–March 2010) and testing period (January–March 2011). (b) Kernel density estimates of RMSE distributions for training and test sets. (c) Violin plots showing the RMSE distribution with overlaid individual points from each of 100 runs for both training and test sets.}
  \label{fig:12}
\end{figure}
To rigorously assess overfitting and generalization in the CL-RCNN and RCNN frameworks, a multi-run analysis using 100 different random seed initializations is conducted. Deep learning models, particularly those applied to spatial climate data, are known to exhibit variability in training behavior due to random weight initialization, data shuffling, and optimizer state. Varying the seed allows us to evaluate how robust the model is to these stochastic processes and to estimate the distribution and variability of model performance, rather than relying on a single run that may not be representative.

\begin{table}[htb]
\centering
\caption{Test RMSE for January--March 2011 across 100 random-seed initializations. Values are reported as the mean and standard deviation across runs.}
\label{tab:seed_variability}
\begin{tabular}{lccc}
\hline
Model & Runs & Mean test RMSE & Standard deviation \\
\hline
RCNN    & 100 & 0.3911 & 0.0321 \\
CL-RCNN & 100 & 0.3581 & 0.0510 \\
\hline
\end{tabular}
\end{table}
Table~\ref{tab:seed_variability} provides an abbreviated comparison of random-seed sensitivity for both models. Across 100 runs, CL-RCNN achieved a
lower mean test RMSE than RCNN ($0.3581$ compared with $0.3911$), corresponding
to an approximately $8.4\%$ reduction. However, CL-RCNN showed slightly greater
run-to-run variability, with a standard deviation of $0.0510$ compared with
$0.0321$ for RCNN.
Therefore, both RCNN and CL-RCNN were trained 100 times using different random
seed initializations on the 2005--2010 training dataset. Table~\ref{tab:seed_variability}
summarizes the test RMSE variability for both models, while Figure \ref{fig:12} provides
a more detailed train--test generalization analysis for CL-RCNN.

Figure \ref{fig:12}(a) shows the daily RMSE, with the solid and dashed lines representing test and train performance, respectively. The shaded bands denote standard deviation. Although the RMSE of the test is consistently higher than the RMSE of the training, the temporal trends remain parallel and the variance between the two is low, indicating stable generalization. To quantify this generalization behavior across seeds, Figure \ref{fig:12}(b) presents kernel density estimates (KDEs) of the RMSE distributions. The training RMSE distribution peaks at 0.316, while the test RMSE peaks at 0.360, indicating a modest generalization gap of approximately 0.044 RMSE units. This gap quantifies the difference in model performance between the data the model has seen during training and unseen data from a different year. A small and consistent shift in the distributional peaks is expected in real-world seasonal prediction tasks, where interannual variability, data heterogeneity, and local anomalies present additional challenges at test time. Additionally, this suggests that the model is learning meaningful spatial-temporal patterns rather than memorizing the training data. 
Figure \ref{fig:12}(c) presents a violin plot that compares the distribution of RMSE values for training and test sets across random seeds.  Both training and testing distributions are compact, with no outliers or skewed tails. The inclusion of individual points highlights the consistency and low variance of the model between random seeds. The test distribution is shifted toward higher RMSE values relative to the training distribution, reflecting the expected generalization gap between seen and unseen data.


Although the multi-seed analysis quantifies sensitivity to random initialization,
it should not be interpreted as a complete estimate of predictive uncertainty.
It captures only one component of epistemic uncertainty and does not account
for uncertainty in the predictor data, observations, model structure, or
irreducible variability. However, the computational efficiency of the proposed
framework allows it to be applied independently to each member of an existing
ensemble forecast, producing an ensemble of high-resolution SST fields from
which forecast spread and threshold-exceedance probabilities can be estimated.
Related probabilistic super-resolution work, such as PODiff
\cite{jadhav2026podiff,jadhav2026patch}, directly generates high-resolution ensembles and
evaluates their uncertainty using coverage, reliability, calibration, and
probabilistic scoring metrics. A comprehensive assessment of ensemble
calibration within the present RCNN framework remains beyond the scope of this
study.

Overall, this analysis shows that the CL-RCNN model does not overfit, maintains consistent performance, and generalizes well to unseen seasonal data. 

\section{Conclusion and Future Work}
In this work, we introduce a residual corrective neural network approach to statistically downscale sea surface temperatures from the global coarse resolution ACCESS-S2 model to a finer regional scale ROMS model. The main findings of this study are:
\begin{itemize}
    \item The RCNN builds on a U-Net architecture and improves predictions through iterative residual refinement, helping to preserve both large-scale ocean patterns and fine-scale coastal features.
    \item The RCNN model performs well for general SST downscaling tasks, capturing spatial patterns, and reducing errors compared to other baseline methods, such as the interpolation model or U-Net.
    \item However, RCNN tends to underestimate extreme SST values, especially during rare events such as marine heatwaves, because of their limited representation in the training dataset.
    \item To better capture marine heatwave events, which are often missing or underrepresented in historical training datasets and whose intensity, frequency, and spatial distribution may change unpredictably in the future, a custom loss-assisted model called CL-RCNN is proposed. This model uses synthetic SST data to guide the learning toward extreme temperature values.
    \item The CL-RCNN model overcame the limitation of lack of temperature extremes in training data by incorporating synthetic MHW-like fields and a custom loss function, allowing the model to better recognize and replicate high-temperature anomalies and improve precision in regions with extreme values (e.g., $>30^\circ$ C).
    \item This analysis highlighted that, for downscaling tasks, especially with strong temporal dynamics, leveraging more recent data not only reduces computational demands but also can yield accurate results. Prioritizing temporally relevant training data ensures that the model captures the most representative spatial structures and variability patterns, making it a practical and effective strategy for data-driven coastal ocean predictions.
    \item Across both generalized cases and the 2011 Western Australian MHW event, RCNN and CL-RCNN significantly outperform standard baseline model like random forest interpolation in terms of RMSE, SSIM, and the ability to reconstruct spatial detail.
    \item The interpretability of the model is enhanced using SHAP analysis, which identifies which physical variables influence the most predictions. Global SST was consistently the most influential input variable aligning with expectations, as it is also the target variable and therefore strongly correlated. Salinity and net longwave radiation followed as the next most important features, reflecting their critical roles.
    \item The developed machine learning framework demonstrates strong potential for flexible and scalable application beyond the specific case of WA marine heatwaves. Its modular design allows for adaptation to different time periods, regions, and climate-related phenomena. This versatility makes it a valuable tool for broader downscaling and prediction tasks in other environmental and oceanographic contexts.
    \item Because CL-RCNN is evaluated only on the 2011 marine heatwave within one Western Australian coastal domain, its transferability to other events and regions remains to be established using independent high-resolution reference datasets.
    \item While the RCNN framework presented in this work performs well in capturing fine-scale SST features and detecting the MHW events, we will explore its generalizability across different regions and oceanographic variables in the future work. The model architecture and training pipeline offer the flexibility to adapt to various spatial resolutions and datasets, but careful tuning of hyperparameters is essential. For example, hyperparameters in the cosine function that governs the smooth refinement over $T$ iterations should be tailored to the specific application and region of interest for optimal performance.
    \item Future work will extend this framework from seasonal predictions to climate projections and include satellite SST comparisons to assess inter-model variability and observational realism.
\end{itemize}

\section*{Open Research Section}
The global climate model sea surface temperature (SST) data used in this study are publicly available from the Australian Bureau of Meteorology \cite{bom2024} and the Copernicus Climate Change Service through the Climate Data Store \cite{hersbach2020era5}. The high-resolution ROMS SST data were obtained from \cite{pattiaratchi2024wamsi}. The Residual Corrective Neural Network (RCNN) framework was developed in Python using TensorFlow and Keras, while the interpolation models were implemented using the Scikit-learn library. The RCNN statistical downscaling framework and supporting scripts used in this study are publicly available via Zenodo (Jadhav, 2026) at https://doi.org/10.5281/zenodo.18349058. The repository is linked to the development GitHub repository and provides a permanently archived, citable version of the code.

\acknowledgments
We acknowledge funding from the State of Western Australia for the project “Advancing predictions of WA Marine Heatwaves and impacts on marine ecosystems". We also thank Dr. C. Spillman (Bureau of Meteorology) for her assistance in accessing and understanding the ACCESS-S2 dataset, which was essential to this study. Dr. Ivica Janekovic, was supported by the WAMSI Westport Marine Science Program during the development of the regional 3D numerical model ROMS used in this work.

\section*{Conflict of Interest}
This work was supported by funding from the State of Western Australia for the project “Advancing predictions of WA Marine Heatwaves and impacts on marine ecosystems". Dr. Janekovic received funding from the WAMSI Westport Marine Science Program for work related to regional ocean modeling. All other authors declare no competing interests.
%
%

\bibliography{agusample}

\appendix
\section{Generalized Case: Sensitivity and Ablation Analyses}
\label{app:general}

\subsection{Sensitivity to Training Period Selection}

This appendix provides a detailed analysis of how the choice of training period influences RCNN performance for the generalized (non-extreme) downscaling case discussed in Section \ref{sec:generalCase}.

We built five different RCNN models using five distinct training periods: 2000–2005, 2005–2010, 2010–2015, 2015–2020, and 2000–2020. Each model was tested on daily SST data for the year 2021. The objective was to examine whether the inclusion of recent years, longer temporal spans, or older historical data affects the model's ability to reproduce SST trends and magnitudes during the test period.

\begin{figure}[htb]
  \centering
  \includegraphics[width=1\columnwidth]{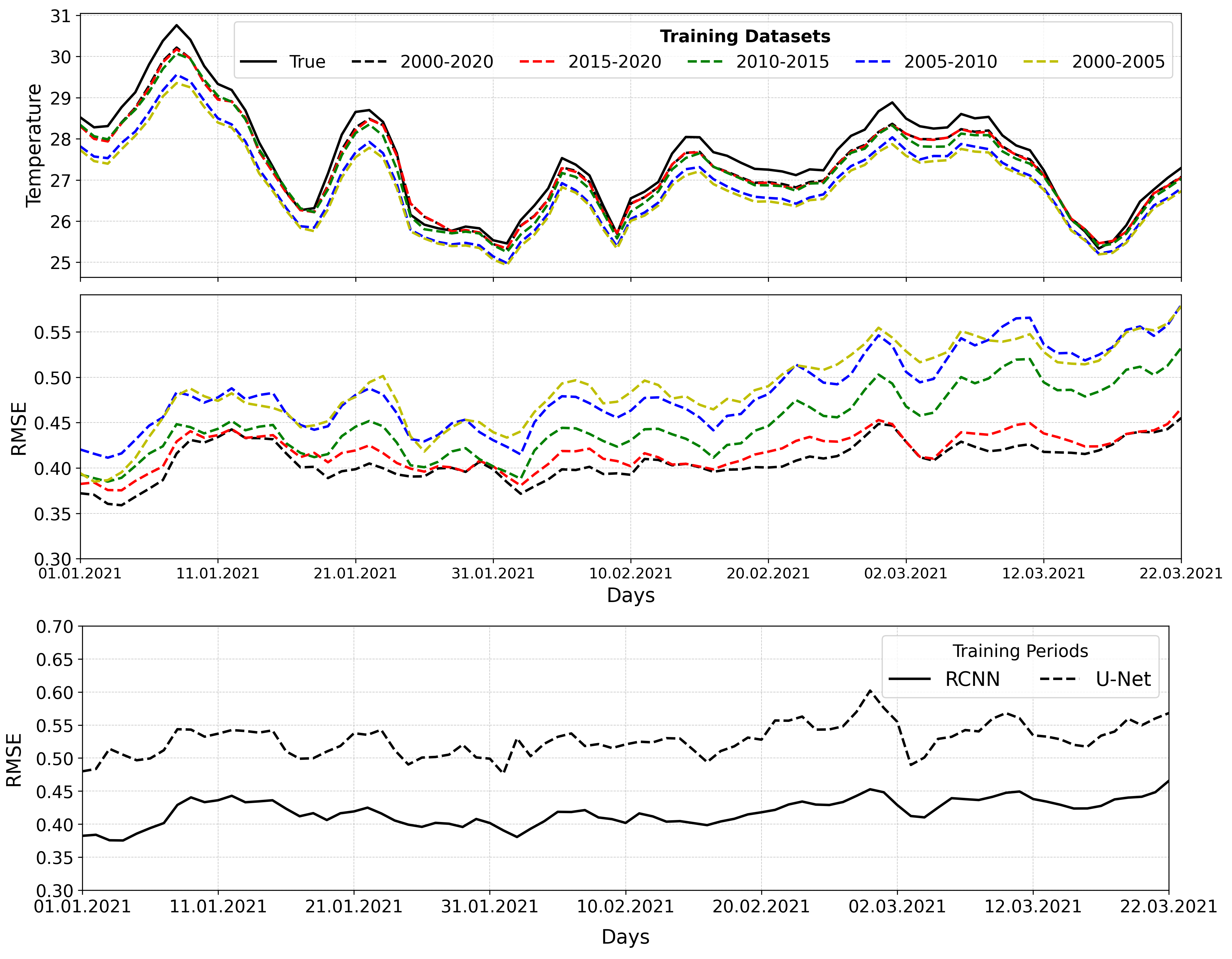}
  \caption{(a) Comparison between the different RCNN models trained on different time period datasets and ROMS based on daily sea surface temperature time series at the coastal location (25.78$^\circ$S, 113.29$^\circ$E) from January to March 2021. (b) Daily Root Mean Squared Error between predicted and true ROMS SST values for each training dataset over the same period for the entire region. (c) Daily RMSE between U-Net and RCNN downscaled SST values for optimal training period (2015-2020) to address the sensitivity of these results to different model architectures.}
  \label{fig:A1}
\end{figure}
Figure \ref{fig:A1}(a) displays the predicted SST time series of each model alongside the true SST (black solid line) at the coastal location (25.78$^\circ$S, 113.29$^\circ$E).
The location is within Shark Bay, which is characterized by complex coastal dynamics, including shallow waters and strong temperature gradients, making it a valuable site for assessing the performance of downscaling models.

Note that the global climate model ACCESS-S2 does not accurately represent this location due to its coarse spatial resolution, which tends to smooth out critical coastal features as seen from Fig. \ref{fig:GlobalvsLocal}. In contrast, our RCNN-based approach enables high-resolution reconstructions that capture these fine-scale dynamics, allowing for more accurate assessments of SST in ecologically and climatically important coastal regions.

Each dashed line in Fig. \ref{fig:A1}(a) represents predictions from a model trained on a different temporal dataset. 
We observed that the model trained on the longest  dataset (2000-2020) performed best due to the diverse representation of SST variability, as expected. 
Furthermore, the model trained on more recent datasets (2015–2020) performed substantially better than models trained on older data windows (2000-2005), closely tracking the true SST fluctuations. The 2015-2020 model predictions capture both the magnitude and timing of temperature peaks and troughs, suggesting good generalization. In contrast, models trained on older data windows (2000–2005 and 2005–2010) tend to systematically underestimate the SST, particularly during peak warming events. This is both reflective of the increasing trend in the SST in this region of the ocean \cite{dalpadado2021warming} and the larger variability of the SST during this period.

Figure \ref{fig:A1}(b) highlights this discrepancy through daily RMSE (Root Mean Squared Error) curves. Lower RMSE values indicate more accurate predictions. Once again, the models trained on 2000–2020 and 2015–2020 consistently yield the lowest RMSE values, typically below 0.45 $^\circ$C, confirming their reliability. The 2000-2020 model often outperforms the rest, highlighting the benefit of exposure to long-term variability and trends in the training data. 
In contrast, models trained on older training sets (2000–2005 and 2005–2010) exhibit the highest RMSE values, sometimes exceeding 0.55 $^\circ$C, especially during mid- to late-February.
These findings suggest that it is important to train your model on data similar to the conditions expected during the prediction or use the custom loss-assisted RCNN to tackle the problem, as we have demonstrated in Subsec. \ref{sec:MHWresults}.
Furthermore, while both the 2000–2020 and 2015–2020 models perform well, it is important to recognize that the 2000–2020 model trains on four times more data, increasing training time, memory usage, and overall computational cost. Therefore, in scenarios where the primary goal is short- to medium-term downscaling, and the system being modeled exhibits clear temporal evolution, prioritizing more recent data can be both computationally efficient and accurate. 
Therefore, unless otherwise stated, all quantitative performance metrics reported in this subsection, including those in Table 3, correspond to models trained on the 2015–2020 dataset and evaluated for the year 2021.

Additionally, to assess the sensitivity of the results to model architecture, we compare the RCNN with a standard U-Net trained using the same 2015–2020 dataset, which represents a good balance between predictive skill and computational cost. This setup isolates architectural effects from those associated with training data selection. The resulting RMSE time series shows that the RCNN consistently outperforms the U-Net throughout the January–March 2021 period in Fig. \ref{fig:A1}(c). This highlights the advantage of the iterative residual correction strategy over a single-pass U-Net model, particularly under dynamically evolving conditions.

\subsection{Ablation Study: Iterative vs One-Step Residual Correction}
This subsection reports an ablation study designed to isolate the effect of iterative residual correction compared to a single-step residual prediction strategy.
\subsubsection{Ablation Study: Iterative vs. One-Step Residual Correction}
To isolate the contribution of iterative refinement, we compare the RCNN against a one-step residual baseline that uses the identical initial U-Net and residual correction architecture but applies the residual prediction only once.
\begin{figure}[htb]
  \centering
  \includegraphics[width=0.8\columnwidth]{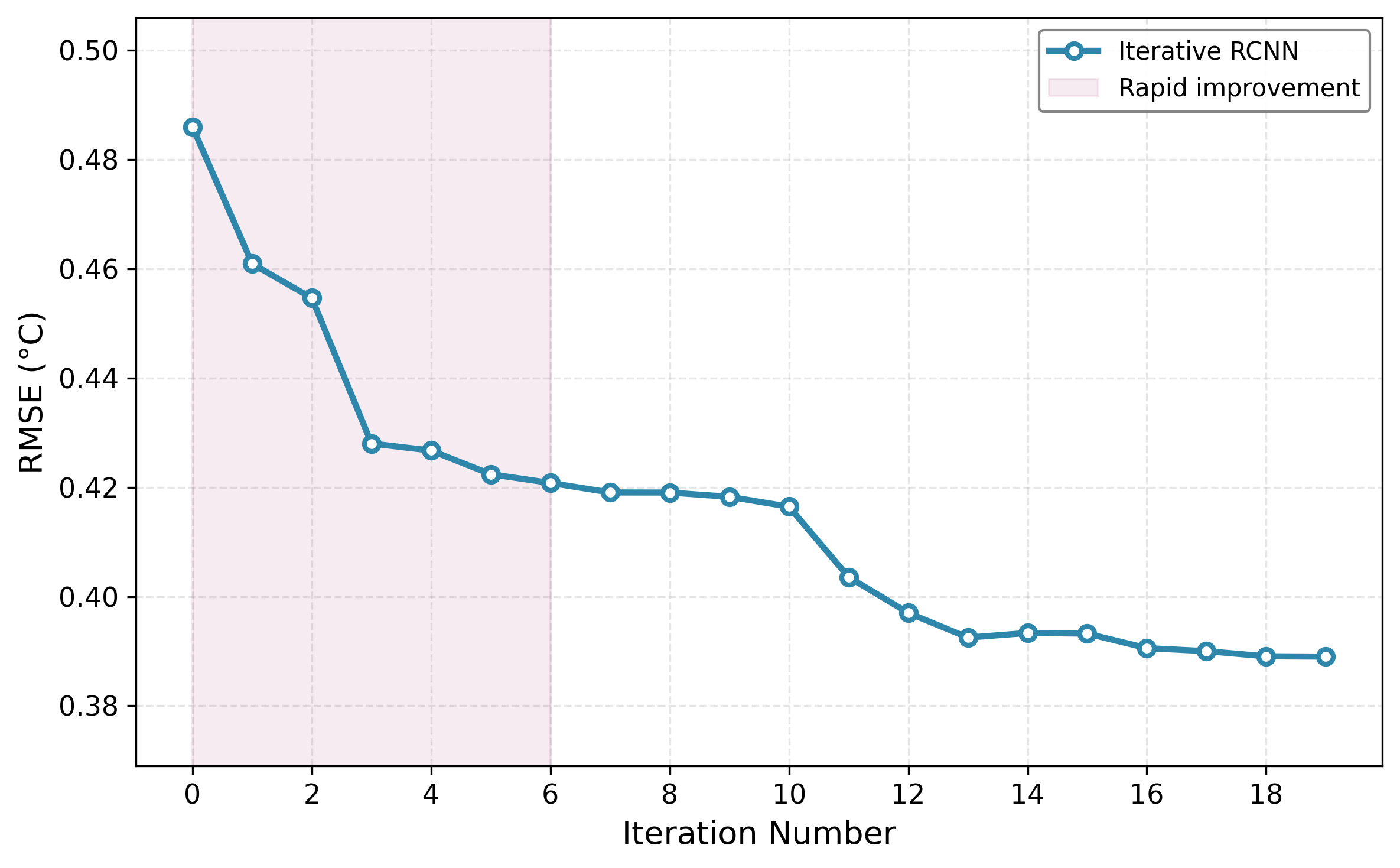}
  \caption{Convergence of the iterative RCNN refinement measured by RMSE as a function of iteration number. RMSE decreases rapidly during the first few iterations, followed by gradual improvement and eventual convergence, indicating stable and effective iterative residual correction.}
  \label{fig:A2}
\end{figure}

Tab. \ref{Tab:ModelPerformance} shows that iterative refinement reduces RMSE by 22.8\% relative to the one-step residual model (0.389 vs 0.504) and improves SSIM from 0.989 to 0.998, demonstrating a clear performance advantage beyond a single residual correction. Additionally, Figure \ref{fig:A2} presents the convergence of the iterative RCNN refinement measured by RMSE as a function of iteration number. Figure shows monotonic convergence of RMSE across iterations quantifying the performance gain or SSIM) of the iterative approach over the one step residual model. The rapid error reduction during the first 5–7 iterations reflects the correction of dominant, large-scale and mesoscale SST biases that are easier for the residual network to learn. Subsequent iterations yield progressively smaller improvements as the model focuses on finer-scale structures and localized gradients, which have lower signal-to-noise ratios and contribute less to global error metrics. The eventual plateau after approximately 15–20 iterations indicates convergence of the iterative refinement process and represents the limits of the model.

\section{Marine Heatwave Case: Training Data Sensitivity}
\label{app:mhw}
This appendix examines the sensitivity of the CL-RCNN model to the choice of training dataset for the marine heatwave case study discussed in Section \ref{sec:MHWresults}.

\subsection{Sensitivity of CL-RCNN to Training Dataset}
\begin{figure}[htb]
  \centering
  \includegraphics[width=0.6\columnwidth]{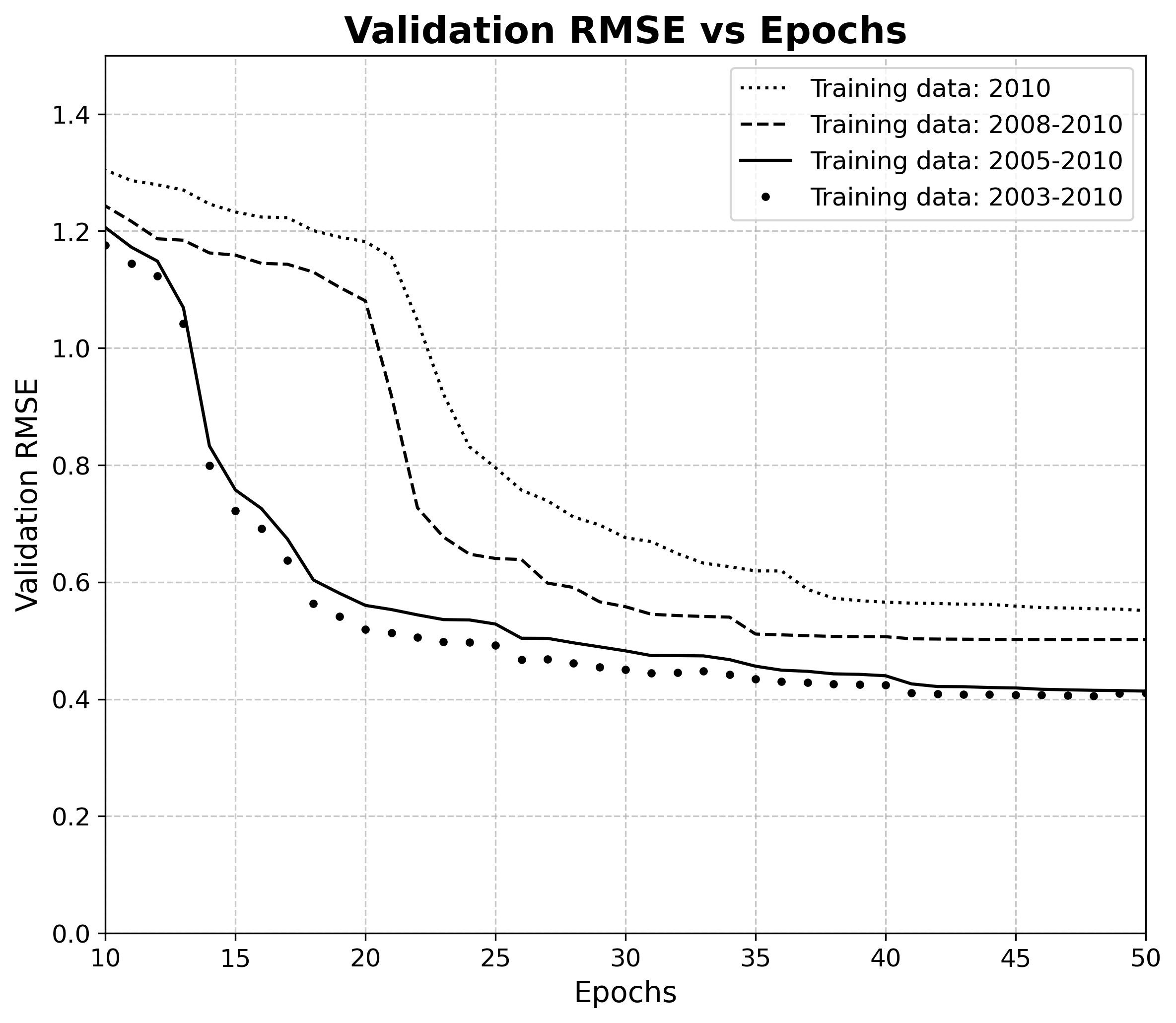}
  \caption{Validation RMSE versus epochs for models trained on different datasets. Although the model is optimized using the custom loss defined in Eq. (3), performance is evaluated using RMSE on the validation set to provide an interpretable error metric. Training periods include 2010 (dotted line), 2008–2010 (dashed line), 2005–2010 (solid line), and 2003–2010 (dotted points).}
  \label{fig:B1}
\end{figure}
To optimize model performance and minimize computational cost, the training dataset is selected dynamically by successively increasing the numbers of years included in the dataset.

Figure \ref{fig:B1} shows the evolution of the validation RMSE estimated over the entire region with respect to the number of training epochs for CL-RCNN models trained on different datasets. Here, epoch means one complete pass through the entire training dataset during model training. The validation is performed based on the last three months of 2010. 
It can be observed that the models trained on larger datasets (e.g., 2003-2010) consistently show lower validation RMSE than the models trained on smaller datasets (e.g., only 2010). This shows that increasing the length of the training dataset improves model performance. By epoch 50, the model trained on the 2003-2010 dataset gives the lowest validation RMSE, which is 0.405 $^\circ$C. However, it is also observed that the use of data from 2005 to 2010 provided results comparable to those obtained with the larger 2003–2010 dataset. The lowest validation RMSE in the case of the 2005-2010 dataset is 0.413 $^\circ$C.
Note that there were notable temperature anomalies toward the end of 2010. To fairly assess the performance of the CL-RCNN model, we withheld the data from October to December 2010 from model training and used this period exclusively for validation.

\begin{table}[htb]
\centering
\caption{Estimated training time for the U-Net model with batch size 8 and 50 epochs on an AMD MI250X GPU. For all configurations, the model input consists of seven predictor channels mapped to the ROMS grid, and the output is a single SST field on the same grid.}
\label{tab:training_time}
\begin{tabular}{lccc}
\hline
Training data& Steps/Epoch& Time/Epoch (min)& Total Time (50 Epochs) (min)\\
\hline
2010& 46   & 1–2& 50–60\\
2008-2010& 137  & 3–3.5& 150-175\\
2005-2010& 274  & 4–4.5& 200-225\\
2003-2010& 365  & 5–5.5& 250-275\\
\hline
\end{tabular}
\end{table}

Furthermore, Tab. \ref{tab:training_time} presents the estimated training time required to train the U-Net model for 50 epochs using a batch size of 8 on a single AMD MI250X GPU. The training datasets span different temporal ranges, increasing in size from a single year (2010) to a multi-year period (2003–2010). As expected, the number of steps per epoch increases proportionally with the number of training samples, leading to longer training times. The per-epoch time ranges account for minor fluctuations due to data loading and computational overhead.

Overall, training with the complete data set from 2003 to 2010 requires the longest compute time (approximately 250-275 min). Furthermore, the RMSE validation results in Fig. \ref{fig:7} show that the model trained on data from 2005 to 2010 gives comparable results with the model trained on 2003-2010 datasets. In short, the 2005–2010 configuration offers a balanced trade-off between predictive performance and computational efficiency, allowing a significant reduction in training time (approximately 50 minutes less) without sacrificing accuracy. Therefore, all the results presented hereon are for the models trained on the 2005-2010 datasets.

\section{Sensitivity of Synthetic Marine Heatwave Generation}
\label{app:synthetic_sensitivity}

This appendix provides additional details on the synthetic marine heatwave data generation procedure used for CL-RCNN training and evaluates the sensitivity of the approach to the synthetic-field construction parameters.

The synthetic MHW fields are generated using time-indexed low-resolution interpolated SST fields rather than a single static field. In this study, $y_{\mathrm{LR}}$ was selected from the interpolated low-resolution SST fields during the January--March 2011 MHW period. The reference high-resolution ROMS SST fields used for injecting the modulated Gaussian features were selected from the first three months of 2010. This choice provides seasonally comparable high-resolution spatial structures while avoiding direct use of the 2011 ROMS target fields during CL-RCNN training.

For each selected time-indexed $y_{\mathrm{LR}}$ field, a corresponding high-resolution ROMS reference field from early 2010 was used to construct one synthetic MHW-like high-resolution SST sample. In total, 90 synthetic MHW-like samples were generated and included during CL-RCNN retraining. These synthetic fields are used only during the offline retraining stage. Once CL-RCNN is trained, end users do not need to generate synthetic fields at inference time. The trained CL-RCNN is applied in the same way as the standard RCNN, using the current prediction and the interpolated low-resolution SST field as inputs.

To assess whether the CL-RCNN improvement depends strongly on one specific synthetic-data design, we performed a sensitivity analysis over the smoothing scale $\sigma_G$ and the injection strength $\gamma$. Three representative settings were considered: a weak/small-scale perturbation setting, the baseline setting used in the main experiments, and a strong/large-scale perturbation setting. For each setting, CL-RCNN was retrained using the corresponding synthetic MHW-like fields and evaluated on the real January--March 2011 MHW period.

\begin{table}[htb]
\caption{Sensitivity of CL-RCNN performance to synthetic MHW generation parameters. The smoothing scale $\sigma_G$ controls the Gaussian background removal in the MHW anomaly extraction, while $\gamma$ controls the strength of synthetic anomaly injection. The RCNN model without synthetic augmentation is included as a reference.}
\label{Tab:SyntheticSensitivity}
\begin{center}
\scriptsize
\begin{tabular*}{\textwidth}{@{\extracolsep{\fill}} l c c c c c}
\hline
\textbf{Model} & $\sigma_G$ & $\gamma$ & \textbf{Jan--Mar} & \textbf{Extreme-pixel} & \textbf{Extreme-day} \\
 & & & \textbf{RMSE} & \textbf{RMSE} & \textbf{RMSE} \\
\hline
RCNN & -- & -- & 0.425 & 0.588 & 0.614 \\
CL-RCNN weak & 3 & 0.3 & 0.369 & 0.478 & 0.520 \\
CL-RCNN baseline & 5 & 0.5 & 0.343 & 0.410 & 0.460 \\
CL-RCNN strong & 10 & 0.7 & 0.430 & 0.562 & 0.590 \\
\hline
\end{tabular*}
\end{center}
\end{table}
Table~\ref{Tab:SyntheticSensitivity} provides a sensitivity analysis of the synthetic MHW generation parameters. The results show that the CL-RCNN improvement is not limited to a single synthetic-data design. The weak/small-scale setting improves over the standard RCNN for all three evaluation metrics, while the baseline setting gives the best overall performance. The strong/large-scale setting remains better than RCNN for the upper-tail and extreme-day metrics, but gives slightly higher Jan--Mar RMSE, indicating that overly strong synthetic perturbations may reduce performance over the broader seasonal distribution. These results suggest that synthetic MHW augmentation is useful for improving extreme-event prediction, but the smoothing scale and injection strength should be tuned using validation data.

%
%
%
%
%

\end{document}